\documentclass[aps,twocolumn,pra,reprint,amsmath,amssymb,floatfix,footinbib,superscriptaddress]{revtex4-1}
\usepackage{float}
\usepackage{amsmath}	
\usepackage{gensymb}
\usepackage{amssymb}
\usepackage{yhmath}
\usepackage{bbold}
\usepackage{hyperref}
\usepackage{tikz}
\usepackage{array}
\hypersetup{colorlinks=true}
\usepackage{upgreek}
\usepackage{graphics}
\usepackage{hyperref}
\usepackage{epsfig}
\usepackage{color}
\usepackage{bm}
\usepackage{graphicx}
\usepackage{indentfirst}
\usepackage{ulem} 
\usepackage{blindtext}
\usepackage[caption=false]{subfig}
\usepackage{natbib}

\usepackage{CJK}        
\usepackage{url}
\usepackage{multirow}
\definecolor{clr_K}{RGB}{128,128,128}
\definecolor{clr_Gamma}{RGB}{127.5,0,127.5}
\definecolor{clr_M1}{RGB}{0,128,0}
\definecolor{clr_M2}{RGB}{255,0,255}
\definecolor{clr_M3}{RGB}{0,128,0}
\definecolor{clr_AFMS1}{RGB}{119,172.5,48}
\definecolor{clr_AFMS2}{RGB}{255,125.5,0}
\definecolor{clr_FMS1}{RGB}{0,229.5,242.2500}
\begin{document}

\title{Successive topological phase transitions in two distinct spin-flop phases on the honeycomb lattice}

\author{Xudong Li}
\affiliation{College of Physics, Nanjing University of Aeronautics and Astronautics, Nanjing 211106, China}
\affiliation{Key Laboratory of Aerospace Information Materials and Physics (NUAA), MIIT, Nanjing 211106, China}

\author{Jize Zhao}
\affiliation{School of Physical Science and Technology $\&$ Key Laboratory of Quantum Theory and Applications of MoE, Lanzhou University, Lanzhou 730000, China}
\affiliation{Lanzhou Center for Theoretical Physics, Key Laboratory of Theoretical Physics of Gansu Province, Lanzhou University, Lanzhou 730000, China}

\author{Jinbin Li}
\email[]{jinbin@nuaa.edu.cn}
\affiliation{College of Physics, Nanjing University of Aeronautics and Astronautics, Nanjing 211106, China}
\affiliation{Key Laboratory of Aerospace Information Materials and Physics (NUAA), MIIT, Nanjing 211106, China}

\author{Qiang Luo}
\email[]{qiangluo@nuaa.edu.cn}
\affiliation{College of Physics, Nanjing University of Aeronautics and Astronautics, Nanjing 211106, China}
\affiliation{Key Laboratory of Aerospace Information Materials and Physics (NUAA), MIIT, Nanjing 211106, China}
\date{\today}

\begin{abstract}
  The Kitaev magnets with bond-dependent interactions have garnered considerable attention in recent years for their ability to harbor exotic phases and nontrivial excitations.
  The topological magnons, which are indicated by nonzero Chern number that can enhance the thermal Hall conductivity, are proposed to partially explain thermal Hall measurements in real materials.
  Hitherto, topological magnons have been extensively explored when the magnetic field is normal to the honeycomb plane, but their topological characteristics are less studied in the presence of in-plane magnetic field.
  Here, we study two distinct in-plane field induced spin-flop phases in the $\Gamma$-$\Gamma'$ model, both of which are off-diagonal couplings that have intimate relation to the Kitaev interaction.
  The two spin-flop phases are distinguished by their out-of-plane spin components which can be either antiparallel or parallel, thus dubbing antiferromagnetic (AFM) or ferromagnetic (FM) spin-flop phases, respectively.
  We map out topological phase diagrams for both phases, revealing a rich pattern of the Chern number over exchange parameters and magnetic field.
  We analytically calculate the boundaries of topological phase transitions when the magnetic field is along the $a$ and $b$ directions.
  We find that the thermal Hall conductivity and its derivative display contrasting behaviors when crossing different topological phase transitions.
  The striking difference of the two phases lies in that when the magnetic field is along the $b$ direction, topological magnons are totally absent in the AFM spin-flop phase, while they can survive in the FM analogue in certain parameter regions.
\end{abstract}

\pacs{}

\maketitle

\section{Introduction}\label{chapter1}
In the last decade, there has been keen interest in frustrated magnets that host topological features of their excitations due to bond-dependent exchanges arising from spin-orbit coupling (for recent reviews, see Refs.~\cite{McClarty2022ARCMP,Xu2022PP,Rousochatzakis2024RoPP,Zhang2024PP}).
Owing to the pioneering work by Kitaev \cite{kitaev2006APNY} and the subsequent materialization via Jackeli-Khaliullin mechanism \cite{Jackeli2009PRL}, the Kitaev interaction, which comprises of bond-directional Ising couplings, emerges as a prominent ingredient for the realization of many alluring phenomena, such as field-induced quantum spin liquids \cite{Baek2017PRL,Zheng2017PRL,Lin2021NC,Li2022PRX,Lin2024arXiv}, topological phase transitions \cite{Gordon2021PRR,Suetsugu2022JPSJ,Hwang2024PRB}, and half-quantization in thermal Hall conductivity or oscillation in longitudinal thermal conductivity \cite{Matsuda2018Nature,Matsuda2021Science,Czajka2021NatPhys,Czajka2023NatMater}.
Nevertheless, the diagonal Kitaev interaction solely does not account for these peculiar experimental results, highlighting the role played by off-diagonal $\Gamma$ and $\Gamma'$ terms \cite{Rau2014PRL,Rau2014arXiv}.
In the existing Kitaev materials like $\alpha$-RuCl$_3$ \cite{Kim2014PRB,HYKee2015PRB,Coldea2015PRB} as well as cobaltates Na$_3$Co$_2$SbO$_6$ and Na$_2$Co$_2$TeO$_6$ \cite{Kim2022JPCM}, the $\Gamma$ interaction is a symmetry-allowed indispensable term as that of the Kitaev interaction and their values are generally comparable \cite{Maksimov2020PRR}.
The $\Gamma'$ interaction, on the other hand, originates from the trigonal distortion of edge-shared octahedra structure in real materials \cite{Rau2014arXiv}.
Whereas its value is tiny or subleading, it is helpful to stabilize the zigzag ordering in the ground state and to clarify the sample dependence of the intermediate region in the magnetic field \cite{Gordon2019NC,Chern2020PRR,Tanaka2020PRB}.
Noteworthily, recent works have revealed that $\Gamma$ and $\Gamma'$ interactions are beneficial for the sought-after quantum spin liquids \cite{Wang2019PRL,Luo2021NPJ,Lee2020NC,Gohlke2020PRR}.
In addition, a novel chiral-spin ordering with spontaneously time-reversal symmetry breaking is identified in the $\Gamma$-$\Gamma'$ model \cite{Luo2022PRR}, and a spin-flop phase that is analogy of the superfluid phase is induced in the presence of magnetic field \cite{Luo2022PRB}.

In addition to the exotic phases, these spin-orbit coupled models also serve as a fertile platform for exploring unconventional excitations like topological magnon \cite{McClarty2022ARCMP}.
Whereas the detection and confirmation of topologically nontrivial magnon bands in Kitaev magnets is still a challenge, theoretical exploration of topological magnons continues gaining momentum.
To commence on, the topological magnons are identified in the fully polarized paramagnetic phase at strong magnetic field.
On the one hand, smoking-gun signatures of topological magnons are verified by the occurrence of nonzero Chern number and chiral edge states in the out-of-plane magnetic field \cite{McClarty2018PRB,Joshi2018PRB}.
On the other hand, in the in-plane magnetic field, the specific sign structures of thermal Hall conductivity along the $a$ and $b$ directions are revealed \cite{Zhang2021PRB,Chern2021PRL}, and the full topological phase diagram with respect to the field angle are studied extensively \cite{Chern2024PRB}.
Turning to the zero or weak magnetic field, it is shown that there are successive topological phase transitions within the parameter region of noncoplanar triple meron crystal \cite{Chen2024PRB}.
In addition, stemming from the trivial antiferromagnetic (AFM) phase, an intermediate spin-flop phase is identified upon increasing the out-of-plane magnetic field \cite{Luo2022PRB,Mertig2022PRL}.
Despite of being different types, they own topological magnons and undergo topological phase transitions as the magnetic field varies.
Quite recently, the spin-flop transition in the Kitaev antiferromagnet Na$_3$Ni$_2$BiO$_6$ \cite{Shangguan2023NP} and Na$_2$Co$_2$TeO$_6$ \cite{Sun2023PRB,Zvyagin2023PRB} are observed experimentally.
However, the topological nature of the spin-flop phase in the in-plane magnetic field is yet less explored.

In this paper, we investigate the topological properties of two distinct spin-flop phases in the $\Gamma$-$\Gamma'$ model subjected to an in-plane magnetic field on a honeycomb lattice \cite{Luo2022PRB}.
Depending on their spin patterns in the out-of-plane direction which are either antiparallel or parallel, they are termed AFM and ferromagnetic (FM) spin-flop phases, respectively.
By using the linear spin-wave theory (LSWT), we map out their topological phase diagrams in the parameter space of exchange couplings and magnetic field at different in-plane field angles.
The topological phase diagrams are primarily occupied by topologically nontrivial region with Chern number $\mathcal{C} = \pm 1$.
The topological phase transitions therein are rather rich, encompassing trivial-nontrivial transition and transition between different nontrivial phases.
We also calculate the magnetic field dependence of the thermal Hall conductivity at selected temperatures.
We find that the thermal Hall conductivity and its first derivative exhibit different behaviors when crossing transition points, indicating that it can serve as a potential probe to capture topological phase transitions.
We report that topological phase transitions in the FM spin-flop phase are richer than that of the AFM spin-flop phase.
In particular, when the magnetic field is along the $b$ direction and other equivalent ones, the Chern number and thermal Hall conductivity in the AFM spin-flop phase is zero entirely.
By contrast, they are largely nonzero and only differ by a minus sign in the two degenerate states of FM spin-flop phase.

This paper is organized as follows.
In Sec.~\ref{chapter2}, we present the $\Gamma$-$\Gamma'$ model under the in-plane magnetic field and outline the LSWT for AFM and FM spin-flop phases conformably.
Subsequent sections detail topological phase transitions within the region of AFM spin-flop phase (Sec.~\ref{chapter3}) and FM spin-flop phase (Sec.~\ref{chapter4}), focusing on the topological phase diagrams via Chern number, thermal Hall conductivity, and chiral edge modes.
Particularly, topological phase transitions in the presence of $a$ and $b$ directional magnetic fields are of major concerns.
In Sec.~\ref{chapter5}, we discuss and conclude our findings.
In the Supplemental Material \cite{SuppMat}, we show topological phase diagrams at three selected in-plane field angles $\varphi$ and provide a $\varphi$-dependence of an animation with one-degree increment for each spin-flop phase.

\section{Model and Method}\label{chapter2}
\subsection{Hamiltonian and classical phase diagram}\label{chapter2sectionA}
\begin{figure}[htbp]
    \centering
    \includegraphics[width=0.95\columnwidth, clip]{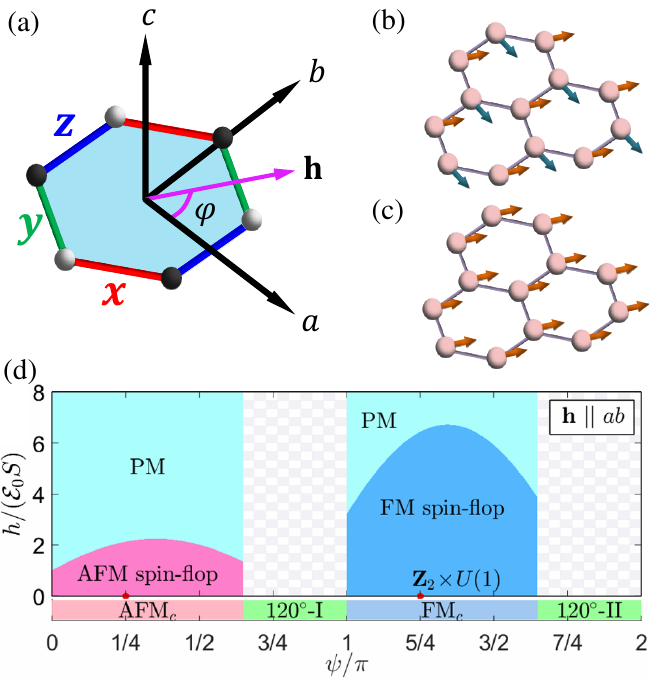}
    \caption{(a) Layout of the honeycomb plane spanned by $a[11\overline{2}]$ and $b[\overline{1}10]$ axes. The $c[111]$ axis is perpendicular to the honeycomb plane. The three types of nearest-neighbor bonds are $x$ (red), $y$ (green), and $z$ (blue) bonds. An external magnetic field $\mathbf{h}$ is applied in the honeycomb plane with an azimuthal angle $\varphi$.
    (b) and (c) show the spin configurations of the coplanar AFM spin-flop and collinear FM spin-flop phases, respectively.
    (d) Classical phase diagram of the $\Gamma$-$\Gamma'$ model under an in-plane magnetic field $\mathbf{h}$, in which $\Gamma = \mathcal{E}_0\cos \psi$ and $\Gamma'= \mathcal{E}_0\sin \psi$. The bottom bar sketches the zero-field phase diagram presented in Ref.~\cite{Luo2022PRB}.
    }\label{Fig:model}
\end{figure}

In strongly spin-orbit coupled materials, their generic spin model with nearest-neighbor interactions is given by \cite{Rousochatzakis2024RoPP,Rau2014PRL}
\begin{align}\label{JKGGpHc-Ham}
\mathcal{H} =
    & \sum_{\left<ij\right>\parallel\gamma} \Big[J \mathbf{S}_i \cdot \mathbf{S}_j + K S_i^{\gamma} S_j^{\gamma}
    + \Gamma \big(S_i^{\alpha}S_j^{\beta}+S_i^{\beta}S_j^{\alpha}\big)\Big]   \nonumber \\
    & + \Gamma' \sum_{\left<ij\right>\parallel\gamma}
        \Big[\big(S_i^{\alpha} + S_i^{\beta}\big) S_j^{\gamma} + S_i^{\gamma} \big(S_j^{\alpha} + S_j^{\beta}\big) \Big]    \nonumber \\
    & - \sum_i \mathbf{h} \cdot \mathbf{S}_i,
\end{align}
where $S_i^{\gamma}$~($\gamma$ = $x$, $y$, and $z$) is the $\gamma$-component of the $\mathbf{S}_i$ in the $xyz$ axis.
Conventionally, the nearest-neighbor bond between site $i$ and $j$ of type $\gamma$ is denoted by $\left<ij\right>\!\parallel\!\gamma$, and $(\alpha, \beta, \gamma)$ is a cyclic permutation of $(x, y, z)$.
In addition, $J$ and $K$ are the diagonal Heisenberg and Kitaev interactions, respectively, while $\Gamma$ and $\Gamma'$ are symmetry-allowed off-diagonal exchanges, and $\mathbf{h}$ represents the magnetic field.
Throughout the paper, we will leave out the diagonal terms, i.e., $J = K = 0$, and exclusively focus on off-diagonal $\Gamma$ and $\Gamma'$ couplings.
Although candidate materials that can be approximated by the minimal $\Gamma$-$\Gamma'$ model are still lacking,
we note that this conception seems to be experimentally feasible since diagonal exchanges are more sensitive to compressive distortions.
It is demonstrated that upon tuning strain in CrSiTe$_3$/CrGeTe$_3$ \cite{Xu2020PRL} or $\alpha$-RuCl$_3$ \cite{Stavropoulos2020unPub}, the Heisenberg interaction or Kitaev interaction varies dramatically, yielding a vanishing value at certain strength of strain.
In the resultant $\Gamma$-$\Gamma'$ model, we parameterize them by using an overall positive energy scale $\mathcal{E}_0$ and an angle $\psi$ such that $(\Gamma,\Gamma') = \mathcal{E}_0(\cos \psi,\sin \psi)$.
For the benefit of the subsequent results, we introduce two auxiliary parameters $\widetilde\Gamma = \Gamma + 2\Gamma'$ and $\overline\Gamma = \Gamma - \Gamma'$.
Meanwhile, we also consider the in-plane magnetic field, which can be parameterized as $\mathbf{h}=h(\cos \varphi, \sin \varphi, 0)$ in the crystallographic $abc$ axis with $a[11\overline{2}]$, $b[\overline{1}10]$, and $c[111]$ [see Fig.~\ref{Fig:model}(a)].

The classical phase diagram of the $\Gamma$-$\Gamma'$ model in the absence of the magnetic field was previously mapped out by the Luttinger-Tisza method \cite{Luo2022PRB}.
It contains four magnetically ordered phases namely the AFM$_c$ phase, the FM$_c$ phase, and the distinct $120^{\circ}$-I and $120^{\circ}$-II phases [see the bottom of Fig.~\ref{Fig:model}(d)].
Of note is that the AFM$_c$ phase is preferred when $\psi\in(0,\psi_0)$ and the FM$_c$ phase is chosen when $\psi\in(\pi,\pi+\psi_0)$, in which $\psi_0 = \pi-\tan^{-1}(2) \approx 0.6476\pi$ and the subscript $c$ denotes that the spins are aligned parallel or antiparallel along the $c$ axis.

When switching on the in-plane magnetic field, parallel tempering Monte Carlo simulations \cite{Metropolis1953,HukushimaNemoto1996} show that the spins in the AFM$_c$ and FM$_c$ phases begin to deviate from the $c$ axis, canting towards the $ab$ plane gradually before enter into the fully polarized paramagnet (PM).
Recalling that the conventional spin-flop phase, which is characterized by a noncollinear spin configuration with two-site unit cell, has a FM order along the longitudinal field direction and displays an AFM order in the transverse plane \cite{Morrison1973PSSB,Fisher1974PRL,Kwon2024PRB}.
Here, we generalize this terminology to accommodate both transverse AFM and FM orderings.
That is, in the AFM (FM) spin-flop phase the spins exhibit an AFM (FM) order in the transverse plane, apart from a FM order along the longitudinal direction.
In this sense, the coplanar AFM spin-flop and the collinear FM spin-flop phases that inherit from the zero-field AFM$_c$ and FM$_c$ phases, respectively, emerge immediately when $h \neq 0$.
We note in passing that the FM spin-flop phase is termed frustrated ferromagnet in some contexts \cite{Vojta2021PRB,Chern2020PRR}.

The natural question to address is the range of the two spin-flop phases.
Since the magnetic field is applied in the honeycomb plane, it is more convenient to use the $abc$ axis.
In this axis, the spin $\mathbf{S}_i$ can be expressed as
\begin{align}\label{EQ:Sthetaphi}
\mathbf{S}_i=S(\sin\theta\cos\phi,\sin\theta\sin\phi,\cos\theta),
\end{align}
where $\theta$ is the polar angle relative to the $c$ axis and $\phi$ is the azimuthal angle in the $ab$ plane.
In the coplanar AFM spin-flop phase, the spins in the two sublattices are given by
$\mathbf{S}_{A} = S(\sin \theta \cos \phi,\sin \theta \sin \phi, \cos \theta )$
and $\mathbf{S}_{B} = S(\sin \theta \cos \phi,\sin \theta \sin \phi,-\cos \theta )$, see Fig.~\ref{Fig:model}(b).
The classical ground-state energy per site $E_\mathrm{cl}=S^2e_\mathrm{cl}$ is given by
\begin{equation}\label{EQ:classicEnergyAFM}
e_\mathrm{cl} = -\frac{\widetilde\Gamma}{2} (2\cos^2\theta + \sin^2\theta) - \frac{h}{S}\sin\theta\cos(\phi-\varphi).
\end{equation}
To minimize $e_\mathrm{cl}$, it is easy to find that $\phi = \varphi$, indicating that the energy becomes independent of the azimuthal angle \cite{Vojta2021PRB}. Thus, $\theta$ is the sole variational parameter and its optimal value is determined by the conditional equations $\mathrm{d} e_\mathrm{cl}/\mathrm{d} \theta = 0$ and $\mathrm{d} e^2_\mathrm{cl}/\mathrm{d} \theta^2 > 0$.
From this we find that
\begin{equation}\label{EQ:PolarAngle}
\theta=\left\{
\begin{array}{lcl}
0,&      & {h=0}\\
\arcsin(\frac{h}{h_c}),&      & {0 < h < h_c}, \\
\frac{\pi}{2},&      & {h\geq h_c}\\
\end{array} \right.
\end{equation}
where the critical field $h_c = S\widetilde\Gamma$ ($\widetilde\Gamma > 0$).
According to the intervals in Eq.~\eqref{EQ:PolarAngle}, the underlying ground states correspond to the $\mathrm{AFM_c}$ phase, AFM spin-flop phase, and PM, respectively, and the classical energy in the AFM spin-flop phase is calculated as $e_\mathrm{cl} = -\widetilde\Gamma-{h^2}/{(2\widetilde\Gamma S^2)}$.
As a comparison, the spins in the collinear FM spin-flop phase are equally of the form $\mathbf{S}_{A, B} = S(\sin \theta \cos \phi,\sin \theta \sin \phi, \cos \theta )$, see Fig.~\ref{Fig:model}(c).
The classical ground-state energy per site is given by
\begin{equation}\label{EQ:classicEnergyFM}
e_\mathrm{cl} = \frac{\widetilde\Gamma}{2} (2\cos^2\theta - \sin^2\theta) - \frac{h}{S}\sin\theta\cos(\phi-\varphi).
\end{equation}
Obviously, the minimization of $e_\mathrm{cl}$ in Eq.~\eqref{EQ:classicEnergyFM} yields $\phi = \varphi$ and the same $\theta$ shown in Eq.~\eqref{EQ:PolarAngle}.
Hence, in the FM spin-flop phase the classical energy $e_\mathrm{cl} = \widetilde\Gamma + {h^2}/{(6\widetilde\Gamma S^2)}$ and the critical field $h_c = -3S\widetilde\Gamma$ ($\widetilde\Gamma < 0$).
Since the classical energy of both AFM spin-flop phase and FM spin-flop phase is independent of in-plane field angle $\varphi$, it is convenient to map out the classical phase diagram of the $\Gamma$-$\Gamma'$ model under an arbitrary in-plane magnetic field $\mathbf{h}$, see Fig.~\ref{Fig:model}(d).
Numerical details pertaining to the Monte Carlo simulation can be found in Sec.~S1 in the Supplemental Material \cite{SuppMat}.
The prefactor 3 in the critical field $h_c$ of the FM spin-flop phase indicates that its area is three times as large as that of the AFM spin-flop phase.
This may hint that topological phase transitions in the former phase are richer than the latter.

\begin{table*}
\caption{\label{Tab-LSWTBdG}
  The terms in the $\mathcal{H}_{\mathbf{k}}$ shown in Eqs.~\eqref{EQ:LSWT} and \eqref{EQ:BdGHam} for the AFM spin-flop and FM spin-flop phases,
  which include the pair ($e_\mathrm{cl}$, $\varepsilon_0$), $\lambda_0(\mathbf{k})$, and $\lambda_1(\mathbf{k})$.
  In the summations for $\lambda_0(\mathbf{k})$ and $\lambda_1(\mathbf{k})$, $\Phi_{\{\upsilon=x,y,z\}}$ takes the values $\{2\pi/3,4\pi/3,0\}$. $\delta_x = (\sqrt{3}/2, 1/2)$, $\delta_y = (-\sqrt{3}/2, 1/2)$, and $\delta_z = (0, -1)$ are unit vectors along the $x$, $y$, and $z$-type bonds, respectively.}
\begin{ruledtabular}
\begin{tabular}{ c c c}
Phases  & Terms    & Expressions        \\
\colrule
\rule{0pt}{1.2em}
& ($e_\mathrm{cl}$, $\varepsilon_0$)
&
$
  \begin{array}{l}
      \left(-\widetilde\Gamma-{h^2}/{(2\widetilde\Gamma S^2)}, 2\widetilde\Gamma\right)
  \end{array}
$
\\
AFM spin-flop
& $\lambda_0(\mathbf{k})$
&
$
  \begin{array}{l}
      \frac{1}{6}\sum_{\upsilon}\Big\{\widetilde\Gamma\sin^2\theta+2\overline\Gamma\big[(\sin^2\theta-2)\cos(2\varphi+\Phi_{\upsilon}) + 2i\cos\theta\sin(2\varphi+\Phi_{\upsilon})\big]\Big\}e^{i\mathbf{k}\bm{\delta}_{\upsilon}}
  \end{array}
$
\\
& $\lambda_1(\mathbf{k})$
&
$
  \begin{array}{l}
      \frac{1}{6}\sum_{\upsilon}\Big\{\widetilde\Gamma(\sin^2\theta+2)+2\overline\Gamma\big[\sin^2\theta\cos(2\varphi+\Phi_{\upsilon}) + \sqrt{2}i\sin\theta\sin(\varphi-\Phi_{\upsilon})\big]\Big\}e^{i\mathbf{k}\bm{\delta}_{\upsilon}}
  \end{array}
$
\\
\colrule
\rule{0pt}{1.2em}
& ($e_\mathrm{cl}$, $\varepsilon_0$)
&
$
  \begin{array}{l}
      \left(\widetilde\Gamma+{h^2}/{(6\widetilde\Gamma S^2)}, -2\widetilde\Gamma\right)
  \end{array}
$
\\
FM spin-flop
& $\lambda_0(\mathbf{k})$
&
$
  \begin{array}{l}
      \frac{1}{6}\sum_{\upsilon}\Big\{\widetilde\Gamma(3\sin^2\theta-2)-\overline\Gamma\big[2\sin^2\theta\cos(2\varphi+\Phi_{\upsilon})
      +\sqrt{2}\sin2\theta\cos(\varphi-\Phi_{\upsilon})\big]\Big\}e^{i\mathbf{k}\bm{\delta}_{\upsilon}}
  \end{array}
$
\\

& $\lambda_1(\mathbf{k})$
&
$
  \begin{array}{l}
      \frac{1}{6}\sum_{\upsilon}\Big\{3\widetilde\Gamma\sin^2\theta + \overline\Gamma\big[2(2-\sin^2\theta)\cos(2\varphi+\Phi_{\upsilon}) - \sqrt{2}\sin2\theta \cos(\varphi-\Phi_{\upsilon}) \\
      \qquad -4i\cos\theta\sin(2\varphi+\Phi_{\upsilon}) +2\sqrt{2}i\sin\theta\sin(\varphi-\Phi_{\upsilon})\big]\Big\}e^{i\mathbf{k}\bm{\delta}_{\upsilon}}
  \end{array}
$
\\
\end{tabular}
\end{ruledtabular}
\end{table*}
\subsection{Linear spin-wave theory}\label{chapter2sectionB}
We employ the LSWT to study the magnon excitations in the AFM spin-flop and FM spin-flop phases.
To apply the Holstein-Primakoff transformation, we begin by rotating the quantization axis of each spin such that it aligns with the spin moments in the local frame, and only keep terms up to quadratic order in the Hamiltonian \cite{Lake2015JPCM,Janssen2019JPCM}.
After applying the Fourier transformation, the Hamiltonian in reciprocal space is given by
\begin{equation}\label{EQ:LSWT}
    \mathcal{H}=NS(S+1)e_\mathrm{cl}+\frac{S}{2}\sum_{\mathbf{k}}\mathrm{\Psi}^{\dagger}_\mathbf{k}\mathcal{H}_{\mathbf{k}}\mathrm{\Psi}_\mathbf{k},
\end{equation}
where $\mathrm{\Psi}_\mathbf{k}=(a_{\mathbf{k}},b_{\mathbf{k}},a^{\dagger}_{-\mathbf{k}},b^{\dagger}_{-\mathbf{k}})^{\mathrm{T}}$ and $\mathcal{H}_{\mathbf{k}}$ is a $2 \times 2$ block Hermitian matrix.
For the two spin-flop phases, $\mathcal{H}_{\mathbf{k}}$ takes the following general form of
\begin{align}\label{EQ:BdGHam}
\mathcal{H}_\mathbf{k} =
	\left(\begin{array}{@{}cc|cc@{}}
	\varepsilon_0           &   \lambda_0(\mathbf{k})           &       0                       & \lambda_1(\mathbf{k})          \\
    \lambda_0^*(\mathbf{k})         &   \varepsilon_0           &   \lambda_1(-\mathbf{k})          & 0                          \\
    \hline
    \rule{0pt}{1.0 em}
    0                           &   \lambda_1^*(-\mathbf{k})        &   \varepsilon_0         & \lambda_0^*(-\mathbf{k})       \\
    \lambda_1^*(\mathbf{k})         &   0                           &   \lambda_0(-\mathbf{k})          & \varepsilon_0
	\end{array}\right),
\end{align}
where the explicit expressions for these terms are shown in Table~\ref{Tab-LSWTBdG}.
$\mathcal{H}_\mathbf{k}$ is diagonalized by the Bogoliubov transformation, $E_{\mathbf{k}}=\mathcal{T}_{\mathbf{k}}^{\dagger}\mathcal{H}_{\mathbf{k}}\mathcal{T}_{\mathbf{k}}$, in which $\mathcal{T}_{\mathbf{k}}$ satisfies the relation $\boldsymbol{\Sigma} = \mathcal{T}_{\mathbf{k}}^{\dagger}\boldsymbol{\Sigma} \mathcal{T}_{\mathbf{k}}$ with $\boldsymbol{\Sigma} = \mathrm{diag}(1, 1, -1, -1)$.
Meanwhile, $E_{\mathbf{k}} = \mathrm{diag}(\omega_{1,\mathbf{k}},\omega_{2,\mathbf{k}},\omega_{1,-\mathbf{k}},\omega_{2,-\mathbf{k}})$ whose diagonal elements are the magnon dispersions $\omega_{n\mathbf{k}}$ of both bands ($n=1, 2$).
By diagonalizing $\boldsymbol{\Sigma}\mathcal{H}_\mathbf{k}$, we obtain magnon dispersions ($\omega_{n\mathbf{k}} = S\chi/2$) from the two positive solutions $\chi$ of the following quartic equation
\begin{equation}\label{EQ:BiQuad}
(\chi^4 - 2p\chi^2 + q) + s(\chi-\varepsilon_0)^2 + t = 0,
\end{equation}
where the coefficients
\begin{subequations}\label{EQ:PQRS}
\begin{align}\label{EQ:pqst}
    p =\,& \varepsilon_0^2 + |\lambda_0(\mathbf{k})|^2-\frac{|\lambda_1(\mathbf{k})|^2+|\lambda_1(-\mathbf{k})|^2}{2},\\
    q =\,& \varepsilon_0^4-\varepsilon_0^2\Big[2|\lambda_0(\mathbf{k})|^2+|\lambda_1(\mathbf{k})|^2 \\ \nonumber
      & +|\lambda_1(-\mathbf{k})|^2\Big]+|\lambda_0(\mathbf{k})^2-\lambda_1^*(\mathbf{k})\lambda_1(-\mathbf{k})|^2,\\
    s =\,& |\lambda_0(\mathbf{k})|^2-|\lambda_0(-\mathbf{k})|^2,\\
    t =\,& -\Big[|\lambda_0(\mathbf{k})|^2-|\lambda_0(-\mathbf{k})|^2\Big]|\lambda_0(\mathbf{k})|^2 \\ \nonumber
      & +2\mathrm{Re}\Big[\lambda_0(\mathbf{k})\big(\lambda_0(\mathbf{k})-\lambda_0^*(-\mathbf{k})\big)\lambda_1(\mathbf{k})\lambda_1^*(-\mathbf{k})\Big].
\end{align}
\end{subequations}

For the FM spin-flop phase, the relation $\lambda_0(\mathbf{k})=\lambda_0^*(-\mathbf{k})$ holds, resulting in $s, t = 0$ \cite{Maksimov2020PRR}.
Thus, Eq.~\eqref{EQ:BiQuad} simplifies to a biquadratic equation, with the magnon dispersions given by
\begin{equation}\label{EQ:DispRelation}
\omega_{1,2}(\mathbf{k})  = \frac{S}{2}\chi = \frac{S}{2}\sqrt{p\pm\sqrt{p^2-q}}.
\end{equation}
Here, $p^2 - q = \Delta_1 + \Delta_2 - \Delta_3$, where
\begin{equation}\label{EQ:Del1Del2}
\left\{
\begin{array}{ll}
  \Delta_1 &= 4\varepsilon_0^2|\lambda_0(\mathbf{k})|^2\\
  \Delta_2 &= \left(|\lambda_1(\mathbf{k})|^2-|\lambda_1(\mathbf{-k})|^2\right)^2/4\\
  \Delta_3 &= \lvert\lambda_0(\mathbf{k})\lambda_1(-\mathbf{k})-\lambda^*_0(\mathbf{k})\lambda_1(\mathbf{k})\rvert^2\\
\end{array}.
\right.
\end{equation}
For the AFM spin-flop phase, the $s$ term does not vanish except at a few high-symmetry points.
Therefore, a concise form of the magnon dispersions is generally unavailable although the exact solution of Eq.~\eqref{EQ:BiQuad} exists in principle.
However, as shown in Sec.~S2 in the Supplemental Material \cite{SuppMat},
it is observed that $\chi$ approaches $\varepsilon_0$ near the topological transition point, and Eq.~\eqref{EQ:DispRelation} provides a good approximation when neglecting the residual term $\mathcal{O} = s(\chi-\varepsilon_0)^2 + t$.

In addition, a topological phase can only exist if $\min_{\mathbf{k}} [\omega_{1\mathbf{k}}] > 0$, and a topological phase transition can only occur when a band-gap vanishes, i.e., $p^2-q = 0$.
Since $\Delta_1$, $\Delta_2$, and $\Delta_3 \ge 0$, the possible solutions can be divided into two categories on the basis of whether $\Delta_1$ is zero or not.
On the one hand, if $\Delta_1 > 0$, it is demonstrated that $\Delta_3 < \Delta_1$ always holds in the region of the two spin-flop phases (see Sec.~S3 in the Supplemental Material \cite{SuppMat}). Therefore, no physical solution exists in this case.
On the other hand, if $\Delta_1 = 0$, which is equivalent to state that $|\lambda_0(\mathbf{k})|^2 = 0$, it is inferred that $\Delta_3$ also vanishes given that $\Delta_3 \leq |\lambda_0(\mathbf{k})|^2(|\lambda_1(\mathbf{k})|+|\lambda_1(-\mathbf{k})|)^2$.
Therefore, the only possibility of band-gap closing is achieved when $\Delta_1 = 0$ and $\Delta_2 = 0$ concurrently.
For the two-band model, the spin-wave energy is given by
\begin{equation}\label{EQ:SW energy}
e_{\mathrm{sw}} = S(S+1)e_{\mathrm{cl}}+\frac{S}{4}\sum_{n=1}^{2}\int\frac{\omega_{n\mathbf{k}}}{(2\pi)^2}d^2 \mathbf{k}.
\end{equation}

Several quantities can be used to capture the topological phase and topological phase transition.
If the $n$-th band is separated from others, the Berry curvature $\Omega_{n\mathbf{k}}$ is given by \cite{PhysRevB.99.054409}
\begin{equation}\label{EQ:Berry Curvature}
\Omega_{n\mathbf{k}}\!=\!-2\mathrm{Im}\!\!\sum^{4}_{\substack{m=1\\ m\neq n}}\!\!\!\frac{ (\boldsymbol{\Sigma}\mathcal{T}^{\dagger}_{\mathbf{k}}\partial_{x}\!\mathcal{H}_\mathbf{k}\mathcal{T}_{\mathbf{k}})_{nm} (\boldsymbol{\Sigma}\mathcal{T}^{\dagger}_{\mathbf{k}}\partial_{y}\!\mathcal{H}_\mathbf{k}\mathcal{T}_{\mathbf{k}})_{mn}}{\big[(\boldsymbol{\Sigma}E_{\mathbf{k}})_{nn}-(\boldsymbol{\Sigma}E_{\mathbf{k}})_{mm}\big]^2}.
\end{equation}
The Chern number of the $n$-th band is defined as
\begin{equation}\label{EQ:Chern number}
    \mathcal{C}_n=\frac{1}{2\pi}\int_{\mathrm{FBZ}} \Omega_{n\mathbf{k}} d^2\mathbf{k},
\end{equation}
where FBZ refers to the first Brillouin zone.
Specifically, for the two-band model, the sum of the Chern numbers equals zero ($\mathcal{C}_1 + \mathcal{C}_2 = 0$)\cite{Zhang2023PRB1,Zhuo2021PRB}, thus, it suffices to focus on the Chern number of the lower band.
Numerically, the Chern number can be precisely calculated numerically using an efficient method based on $U(1)$ lattice gauge theory \cite{Suzuki2005JPSJ,Chern2024PRB}.
The thermal Hall conductivity in the $ab$ plane is given by \cite{PhysRevLett.106.197202,PhysRevB.89.054420,MurakamiJPSJ2017,Koyama2021PRB}
\begin{equation}\label{EQ:KXY}
    \kappa_{ab}=-\frac{k_\mathrm{B}^2T}{4\pi^2\hbar}\sum^{2}_{n=1}\sum_{\mathbf{k}\in \mathrm{FBZ}}c_2\big[g(\omega_{n\mathbf{k}})\big]\Omega_{n\mathbf{k}},
\end{equation}
where $g(x)$ is the Bose-Einstein distribution and $c_2(x)=(1+x)\mathrm{ln^2}\big[(1+x)/x\big]-\mathrm{ln^2}x-2\mathrm{Li}_{2}(-x)$, where $\mathrm{Li}_{2}(x)$ is the polylogarithmic function.

\section{AFM spin-flop phase}\label{chapter3}

In this section, we investigate the topological phase transitions within the AFM spin-flop phase under an in-plane magnetic field.
We first map out the topological phase diagram when $\mathbf{h}\,||\,a [11\bar2]$ in Sec.~\ref{chapter3sectionA}.
Next, a theorem that guarantees the Chern number is zero when $\mathbf{h}\,||\,b [\bar110]$ is studied in Sec.~\ref{chapter3sectionB}.
Sec.~\ref{chapter3sectionC} discusses the topological phase transitions in a general direction.
\subsection{Topological phase transitions when $\mathbf{h}\,||\,a [11\bar2]$}\label{chapter3sectionA}
We begin by studying the topological phase transitions when $\mathbf{h}\,||\,a$.
When $h = 0$, the underlying AFM$_c$ phase has nonreciprocal magnons since the energy at the two inequivalent $\textbf{K}$ and $\textbf{K}'$ points is unequal \cite{Luo2022PRB,Satoru2020PRB}.
The two bands touch at the high-symmetry points $\boldsymbol{\Gamma}$ and $\textbf{K}'$.
Specifically, when $\psi/\pi = 0.25$ (i.e., $\overline{\Gamma}=0$), the model reduces to the easy-axis XXZ model \cite{Giniyat2015PRB}, the two bands become completely degenerate and the band-gap does not opens even under the magnetic field.
Thus, this parameter is excluded from the current study.
When $0 < h < h_c$, the system enters the gapped AFM spin-flop phase.
In this region, the band-gap begins to open as long as the magnetic field is applied and it undergoes intricate behaviors due to the interplay of magnetic field and exchange parameters.
Noteworthily, the closing of the band-gap serves as a potential signature of topological phase transition.
Finally, the ground state transitions to the PM when $h > h_c$, and the analytical topological phase diagram of this part was recently proposed by Chern and Castelnovo \cite{Chern2024PRB}.

\begin{figure}[htbp]
    \centering
    \includegraphics[width=0.95\linewidth]{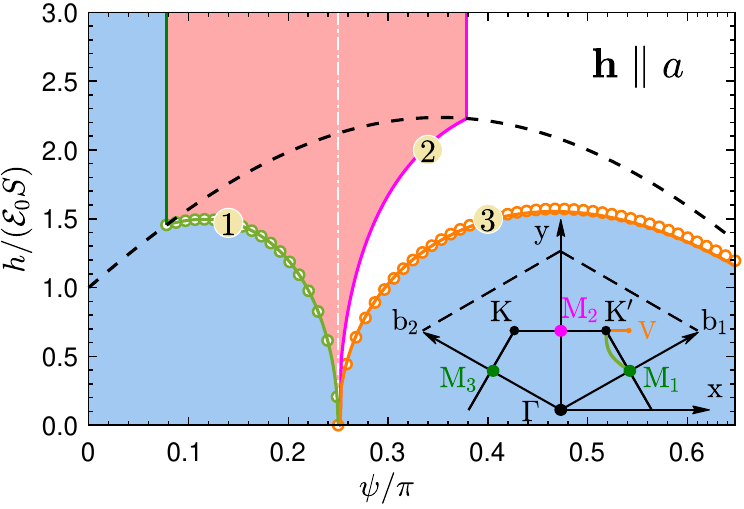}
    \caption{Topological phase diagram of the AFM spin-flop phase under an $a$-directional magnetic field.
    The red, white, and blue areas indicate $\mathcal{C}_1$ = $+1$, $0$, and $-1$, respectively.
    The black dashed curve denotes the magnetic phase transition to the PM.
    In line 2 the curve marks the analytical topological phase boundary.
    In lines 1 and 3, the symbols represent the numerical topological phase boundaries while the curves mark the approximate topological phase boundaries.
    The band-gap closing points and paths are labeled in the inset with the same color.
    Inset: the rhombic Brillouin zone with the relevant high-symmetry points.}
    \label{Fig:AFM diagram}
\end{figure}

To determine the entire topological properties, we have calculated the Chern number $\mathcal{C}_1$ throughout the parameter space and the resultant topological phase diagram is shown in Fig.~\ref{Fig:AFM diagram}.
Originating from the hidden $\mathbb{Z}_2 \times U(1)$ symmetric point at $\psi/\pi = 0.25$, it is observed that there are regions of $\mathcal{C}_1 = -1$ (in blue), $\mathcal{C}_1 = + 1$ (in red), and $\mathcal{C}_1 = 0$ (in white).
In the following sections, we will study the band-gap closing points and the topological phase boundaries based on Eq.~\eqref{EQ:DispRelation}.
In the honeycomb lattice, the primitive lattice vectors are $\mathbf{a}_1 = \sqrt{3}[\cos(\pi/3)\hat{\mathbf{x}}+\sin(\pi/3)\hat{\mathbf{y}}]$ and $\mathbf{a}_2 = \sqrt{3}[\cos(2\pi/3)\hat{\mathbf{x}}+\sin(2\pi/3)\hat{\mathbf{y}}]$, where $\hat{\mathbf{x}}$ and $\hat{\mathbf{y}}$ represent unit vectors along the $a$ and $b$ directions, respectively.
The reciprocal lattice vectors are $\mathbf{b}_1 = 2\pi \hat{\mathbf{b}}_1$ and $\mathbf{b}_2 = 2\pi \hat{\mathbf{b}}_2$,
where $\hat{\mathbf{b}}_1=2/3[\cos(\pi/6)\hat{\mathbf{x}}+\sin(\pi/6)\hat{\mathbf{y}}]$ and $\hat{\mathbf{b}}_2 = 2/3[\cos(5\pi/6)\hat{\mathbf{x}}+\sin(5\pi/6)\hat{\mathbf{y}}]$.
The first Brillouin zone is equivalent to the rhombic region $\mathbf{k} = k_1\hat{\mathbf{b}}_1+k_2\hat{\mathbf{b}}_2$
with $k_{1,2} \in [0,2\pi)$ in reciprocal space (see the inset of Fig.~\ref{Fig:AFM diagram}).
On the other hand, if we write $\mathbf{k} = k_x\hat{\mathbf{x}}+k_y\hat{\mathbf{y}}$, then $k_x, k_y$ and $k_1, k_2$ are related by a linear transformation
\begin{align}
\begin{pmatrix}
k_x\\
k_y
\end{pmatrix} =\frac{2}{3}\begin{pmatrix}
	\cos(\pi/6)          &  \cos(5\pi/6)\\
    \sin(\pi/6)         &   \sin(5\pi/6)\\
	\end{pmatrix}\begin{pmatrix}
	k_1         \\
    k_2         \\
	\end{pmatrix}.
\end{align}

To make out the topological phase transitions, it is helpful and constructive to uncover the conditions for the closing of band-gap \cite{Chern2024PRB}.
For convenience, we introduce six real numbers which are functions of the spin angle $\theta \in (0,\pi/2)$, i.e.,
$c_1 = \widetilde\Gamma\sin^2\theta-\overline\Gamma(\sin^2\theta-2)$,
$c_2 = 3\overline\Gamma(\sin^2\theta-2)$,
$c_3 = 2\sqrt{3}\,\overline\Gamma\cos\theta$,
$c_4 = \widetilde\Gamma(\sin^2\theta+2)-\overline\Gamma\sin^2\theta$,
$c_5 = 3\overline\Gamma\sin^2\theta$, and
$c_6 = -\sqrt{6}\,\overline\Gamma\sin\theta$.
We note that $\widetilde\Gamma$ and $\overline\Gamma$, as well as $c_3$ and $c_6$, are nonzero numbers.
We also bring in two auxiliary functions $f_{\mathbf{k}} = 1+e^{ik_1}+e^{ik_2}$ and $g_{\mathbf{k}} = e^{ik_1}-e^{ik_2}$.
Thus, $\lambda_0(\mathbf{k})$ and $\lambda_1(\mathbf{k})$ with $\varphi=0$ in Table~\ref{Tab-LSWTBdG} are simplified to
\begin{subequations}
\begin{align}
\lambda_0(\mathbf{k}) = \frac{1}{6}(c_1f_{\mathbf{k}}^*+c_2+c_3ig_{\mathbf{k}}^*), \label{EQ:AFMFldALmabda0} \\
\lambda_1(\mathbf{k}) = \frac{1}{6}(c_4f_{\mathbf{k}}^*+c_5+c_6ig_{\mathbf{k}}^*). \label{EQ:AFMFldALmabda1}
\end{align}
\end{subequations}
By using of Eq.~\eqref{EQ:AFMFldALmabda1}, $\Delta_2$ in Eq.~\eqref{EQ:Del1Del2} turns to
\begin{align}\label{EQ:Delta2_reduced_AFM}
\Delta_2& = \left(\frac{c_6}{18}\right)^2\Big[c_4\mathrm{Im}(f^*_{\mathbf{k}}g_{\mathbf{k}}) + c_5\mathrm{Im}(g_{\mathbf{k}})\Big]^2.
\end{align}

Next, we consider $\Delta_1 = 0$, which is equivalent to say that $\left|\lambda_0(\mathbf{k})\right| = 0$ since $\varepsilon_0 \neq 0$ in the parameter region.
In this sense, the real and imaginary parts of $\lambda_0(\mathbf{k})$ should be zero, i.e.,
\begin{subequations}\label{EQ:AFMFldALambda0RealImag}
\begin{align}
&c_1(1+\cos k_1+\cos k_2) + c_2 = c_3(\sin k_1-\sin k_2),\label{EQ:AFMReal}\\
&c_1(\sin k_1+\sin k_2)+c_3(\cos k_1-\cos k_2)=0.\label{EQ:AFMImag}
\end{align}
\end{subequations}
In addition, in light of Eq.~\eqref{EQ:Delta2_reduced_AFM}, $\Delta_2 = 0$ implies that
\begin{align}\label{EQ:AFMFldALambda1}
\left(\!\frac{c_6}{18}\!\right)^2\Big[\!(c_4\!+\!c_5)(\sin k_1\!\!-\!\sin k_2)\!+\!2c_4\sin(k_1\!\!-\!k_2)\!\Big]^2\!\!\!\!=\!0.
\end{align}
There are a couple of possibilities for the solutions of Eqs.~\eqref{EQ:AFMFldALambda0RealImag} and \eqref{EQ:AFMFldALambda1}.
Below we will present these conditions on a case-by-case basis.
The first category is based on the condition $\sin k_1+\sin k_2 \neq 0$.
In this case, the Eqs.~\eqref{EQ:AFMReal} and \eqref{EQ:AFMImag} together lead to $c_1 + c_2 = 0$.
Substituting it into Eq.~\eqref{EQ:AFMFldALambda1}, one finds that
\begin{subequations}\label{EQ:K1K2AFM}
\begin{align}
&\cos\Big(\frac{k_1-k_2}{2}\Big)=\frac{c_3}{\sqrt{c_1^2 + c_3^2}},\label{EQ:K1_K2}\\
&\cos\Big(\frac{k_1+k_2}{2}\Big)=-\frac{2c_4}{c_4+c_5}\cos\Big(\frac{k_1-k_2}{2}\Big).\label{EQ:K1K2}
\end{align}
\end{subequations}
Here, the position of band-gap closing point shifts from $\textbf{K}'$ to $\textbf{M}_1$ as parameter $\psi$ decreases (see the curve in inset of Fig.~\ref{Fig:AFM diagram}).
By contrast, the second and third categories are based on the condition $\sin k_1+\sin k_2 = 0$.
With Eq.~\eqref{EQ:AFMImag} in mind, it is found that $\cos k_1 = \cos k_2$, yielding $k_2 = 2\pi - k_1$.
As a result, Eq.~\eqref{EQ:AFMReal} and Eq.~\eqref{EQ:AFMFldALambda1} are reduced to
\begin{subequations}\label{EQ:AFMdelta12v2}
\begin{align}
&c_1(1 + 2\cos k_1) + c_2 - 2c_3\sin k_1 = 0,\label{EQ:delta1realv2}\\
&\left(\frac{c_6}{9}\right)^2\Big[c_4(1+2\cos k_1) + c_5\Big]^2\sin^2k_1 = 0.\label{EQ:deltav2}
\end{align}
\end{subequations}
Consequently, the second category relies on $\sin k_1 = 0$, from which we can get $k_1$ = 0 or $\pi$.
When $k_1 = 0$, it is inferred from Eq.~\eqref{EQ:delta1realv2} that $3c_1 + c_2 = 0$ or equivalently $\widetilde\Gamma\sin^2\theta = 0$, which has no solution in the range of $\theta\in (0,\pi/2)$.
Thus, the physical solution is $k_1 = \pi$, leading to the constraint equation $c_1 - c_2 = 0$.
The third category relies on $c_4(1+2\cos k_1) + c_5=0$. From Eq.~\eqref{EQ:delta1realv2}, it follows that $(c_2c_4 - c_1c_5)^2 - c_3^2(3c_4^2 - 2c_4c_5 - c_5^2) = 0$.
Then the corresponding topological phase boundary can be obtained subsequently.

The loci of the band-gap closing points in the reciprocal space and the (approximate) topological phase boundaries are plotted in Fig.~\ref{Fig:AFM diagram} and summarized in Table~\ref{Tab:AFMFldADir}.
For line 2, in which the band-gap closing point locates at the $\mathbf{M}_2$ point with $(k_1, k_2) = (\pi,\pi)$, it is calculated that $f_{\bf k} = -1$ and $g_{\bf k} = 0$, and thus there exists $\lambda_0(\mathbf{k}) = \lambda_0^*(-\mathbf{k})$.
Since $s$ and $t$ in Eq.~\eqref{EQ:PQRS} vanish accordingly, this ensures line 2 be an exact topological phase boundary.
By contrast, for line 1 and line 3, which occupy opposite sides of $\psi = \pi/4$, the relation $\lambda_0(\mathbf{k}) = \lambda_0^*(-\mathbf{k})$ does not hold any more at band-gap closing point, giving rise to a nonzero residual term.
Considering the band-gap closing conditions $\Delta_1 = 0$ and $\Delta_2 = 0$, which mean that $|\lambda_0(\mathbf{k})| = 0$ and $|\lambda_1(\mathbf{k})| = |\lambda_1(-\mathbf{k})|$, the residual term simplifies to
\begin{equation}\label{EQ:ResidualTermLine13}
\mathcal{O} = -|\lambda_0(-\mathbf{k})|^2\big(\sqrt{\varepsilon_0^2-|\lambda_1(\mathbf{k})|^2}-\varepsilon_0\big)^2.
\end{equation}
In the spirit of Newton iteration method, value of $\mathcal{O}$ determines the precision of trial solution in Eq.~\eqref{EQ:DispRelation}.
It is surprise to find that $\mathcal{O}$ is reasonably small when compared with the energy at which the two magnon bands touch
(see Fig.~S3 in the Supplemental Material \cite{SuppMat}).
Since the approximation does not alter the dispersion energy too much,
it is foreseeable that there is no apparent difference between the approximate expressions and the exact calculations, see Fig.~\ref{Fig:AFM diagram}.

\begin{table}[th!]
\caption{\label{Tab:AFMFldADir}
    Summary of all band-gap closing scenarios for the AFM spin-flop phase in the $a$-directional magnetic field.
    It includes the band-gap closing points, the corresponding colors drawn in Fig.~\ref{Fig:AFM diagram}, and the expressions of the topological phase boundaries.}
    \begin{ruledtabular}
    \begin{tabular}{c c c c}
    Lines &Points& Colors & Expressions\\
    \colrule
    \rule{0pt}{1.5em}
    \!\!1 &$\wideparen{\textbf{K}'\textbf{M}_1}$& \begin{tikzpicture}    %
    \draw (0,0)--(1.2,0)[color=clr_AFMS1][line width=4pt];
    \end{tikzpicture}
    & $h \simeq 2S\widetilde\Gamma\sqrt{\frac{\overline\Gamma}{\widetilde\Gamma+2\overline\Gamma}}$ \\
    2 &$\textbf{M}_2$& \begin{tikzpicture}
    \draw (0,0)--(1.2,0)[color=clr_M2][line width=4pt];
    \end{tikzpicture}
    & $h = 2S\widetilde\Gamma\sqrt{\frac{2\overline\Gamma}{4\overline\Gamma-\widetilde\Gamma}}$ \\
    3 &$\overline{\textbf{K}'\textbf{V}}$& \begin{tikzpicture}
    \draw (0,0)--(1.2,0)[color=clr_AFMS2][line width=4pt];
    \end{tikzpicture}
    & $h\simeq S\widetilde\Gamma\sqrt{\frac{3\widetilde\Gamma-4\overline\Gamma-\sqrt{9\widetilde\Gamma^2-56\widetilde\Gamma\overline\Gamma+144\overline\Gamma^2}}{8\overline\Gamma-2\widetilde\Gamma}}$\\
\end{tabular}
\end{ruledtabular}
\end{table}

\begin{figure*}[htbp]
    \centering
    \includegraphics[width=0.95\linewidth]{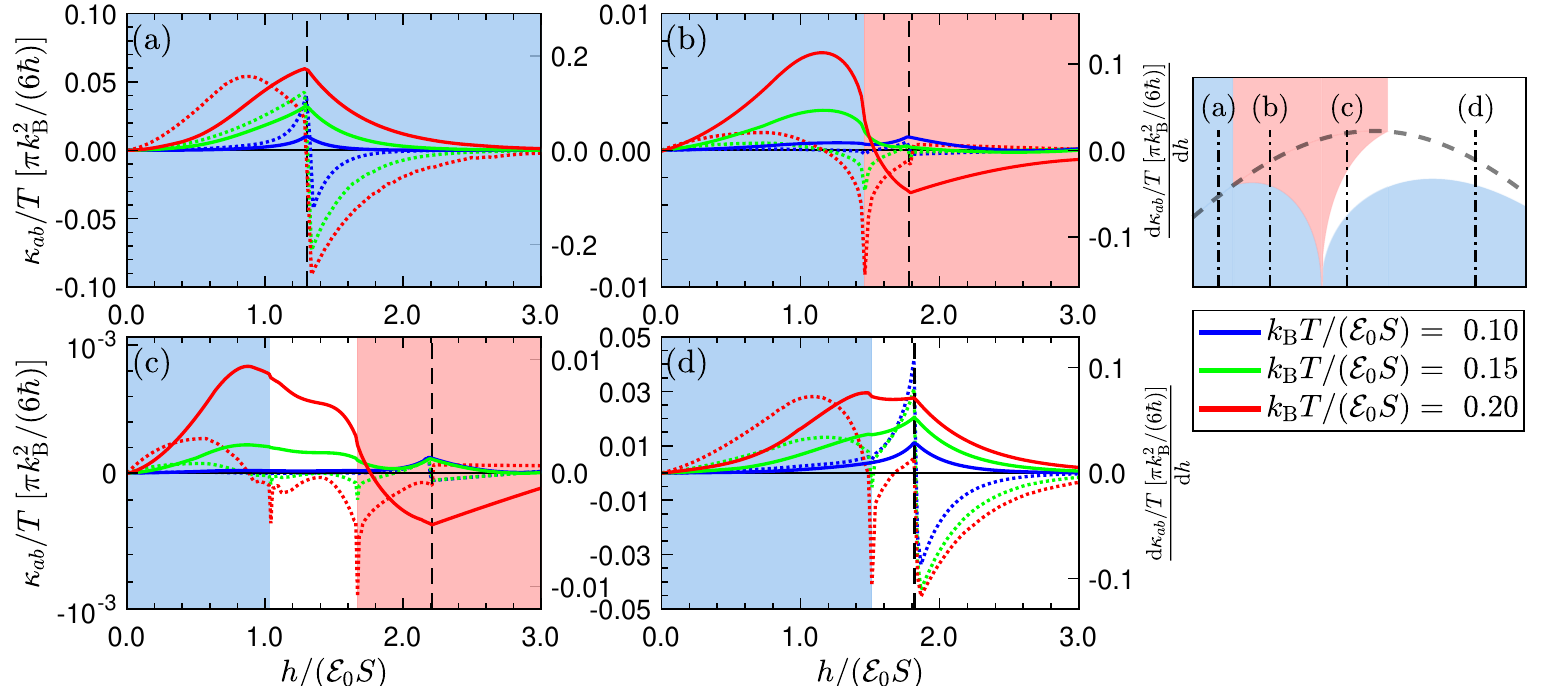}\\
    \caption{Thermal Hall conductivity $\kappa_{ab}/T$ [in units of $\pi k^2_\mathrm{B}/(6\hbar)$] (solid curves) and its first derivative $\mathrm{d}(\kappa_{ab}/T)/\mathrm{d} h$ (dotted curves) vs magnetic field $h$ in the parameter of $\psi/\pi = $ (a) $0.05$, (b) $0.15$, (c) $0.30$, and (d) $0.55$.
    The magnetic field is along the $a$ direction and spin $S = 1/2$ is used.
    The red, white, and blue areas indicate $\mathcal{C}_1$ = $+1$, $0$, and $-1$, respectively.
    The black dashed curve denotes the magnetic phase transition to the PM.
    The rightmost panel shows the position of the parameters in the topological phase diagram.
    }\label{Fig:AFM kxy}
\end{figure*}

In the end, we discuss the relation between thermal Hall conductivity and different types of phase transitions.
Figure~\ref{Fig:AFM kxy} shows $\kappa_{ab}/T$ and its first derivative as a function of magnetic field at temperatures $k_BT/(\mathcal{E}_0S)$ = 0.1 (blue), 0.15 (green), and 0.2 (red).
The four panels (a)-(d) correspond to $\psi/\pi$ = 0.05, 0.15, 0.30, and 0.55, respectively (see the rightmost panel of Fig.~\ref{Fig:AFM kxy}).
Firstly, as can be seen from Fig.~\ref{Fig:AFM kxy}(a) and other panels at high field, $\kappa_{ab}/T$ is continuous albeit with a kink near the magnetic phase transition that is marked by the black dashed line.
In parallel, the derivative of $\kappa_{ab}/T$ has a jump close to the phase transition point.
Secondly, pertaining to the topological phase transition from the topological phase to the trivial, $\kappa_{ab}/T$ displays a jump which becomes clearer with the increase of temperature, see Figs.~\ref{Fig:AFM kxy}(c,d).
Finally, while $\kappa_{ab}/T$ is smoothly varied when crossing the topological phase transition between two distinct topological phases [see Fig.~\ref{Fig:AFM kxy}(b)], it undergoes a sign change at moderate temperature, in accordance with the sign change of the Chern number $\mathcal{C}_1$.
Further, there is a sharp peak in its derivative, indicating the singularity of $\mathrm{d}(\kappa_{ab}/T)/\mathrm{d}h$.
It is worth noting that $\mathrm{d}(\kappa_{ab}/T)/\mathrm{d}h$ is demonstrated to display a logarithmic divergence at the transition point in a triangular-lattice antiferromagnet \cite{Park2019PRB}.

\subsection{Topological phase transitions when $\mathbf{h}\,||\,b [\bar110]$}\label{chapter3sectionB}

Due to the sublattice symmetry, the two degenerate states of the AFM spin-flop phase are equivalent.
This property renders the system to obey the following theorem when the magnetic field is along the $b$ direction.

\textbf{Theorem 1.} Chern number $\mathcal{C}_n$ and thermal Hall conductivity $\kappa_{ab}$ in the AFM spin-flop phase with the magnetic field along $b$ direction is zero.

\textit{Proof}. First, we define two points $\xi=(k_x,k_y)$ and $\eta=(k_x,-k_y)$, which are symmetric about the $x$ axis in reciprocal space, and introduce an orthogonal transformation $U$:
\begin{equation}\label{EQ:U}
U=
\begin{pmatrix}
0&1&0&0\\
1&0&0&0\\
0&0&0&1\\
0&0&1&0\\
\end{pmatrix}
=\mathbb{1}_{2\times2}\otimes\sigma_x,
\end{equation}
where $\sigma_x$ is the Pauli matrix.
By setting $\varphi = \pi/2$ in Table~\ref{Tab-LSWTBdG}, we can obtain the relation of $\mathcal{H}_{\xi}$ and $\mathcal{H}_{\eta}$
\begin{equation}\label{EQ:Hdirb}
\mathcal{H}_{\xi}=U^{\dagger} \mathcal{H}_{\eta}U,
\end{equation}
and their derivative satisfies
\begin{equation}\label{EQ:Hdxydirb}
\frac{\partial \mathcal{H}_{\xi}}{\partial k_x}=U^{\dagger}\frac{\partial \mathcal{H}_{\eta}}{\partial k_x}U,~~\frac{\partial \mathcal{H}_{\xi}}{\partial k_y}=-U^{\dagger}\frac{\partial \mathcal{H}_{\eta}}{\partial k_y}U.
\end{equation}
Suppose that $\mathcal{T}_{\xi}$ is the Bogoliubov transformation of $\mathcal{H}_{\xi}$, then it can be directly deduced that $E_{\xi}=\mathcal{T}_{\xi}^{\dagger}(U^{\dagger}\mathcal{H}_{\eta}U)\mathcal{T}_{\xi}=(U\mathcal{T}_{\xi})^{\dagger}\mathcal{H}_{\eta}(U\mathcal{T}_{\xi})=E_{\eta}$ and $\mathcal{T}_{\xi}^{\dagger}\boldsymbol{\Sigma}\mathcal{T}_{\xi}=(U\mathcal{T}_{\xi})^{\dagger}\boldsymbol{\Sigma}(U\mathcal{T}_{\xi})=\boldsymbol{\Sigma}$.
Thus, we infer that $U\mathcal{T}_{\xi}$ is the Bogoliubov transformation of $\mathcal{H}_{\eta}$.
Meanwhile, in light of Eqs.~\eqref{EQ:Hdirb} and \eqref{EQ:Hdxydirb}, the Berry curvatures at $\xi$ and $\eta$ are related by
\begin{align}
\Omega_{n\xi}=&-2\mathrm{Im}\sum^{4}_{\substack{m=1\\ m\neq n}}\frac{ (\boldsymbol{\Sigma}\mathcal{T}^{\dagger}_{\xi}\partial_{k_x}\mathcal{H}_{\xi}\mathcal{T}_{\xi})_{nm} (x\rightarrow y,m\leftrightarrow n)}{[(\boldsymbol{\Sigma}E_{\xi})_{nn}-(\boldsymbol{\Sigma}E_{\xi})_{mm}]^2}\nonumber\\
=&2\mathrm{Im}\sum^{4}_{\substack{m=1\\ m\neq n}}\frac{ (\boldsymbol{\Sigma}\mathcal{T}^{\dagger}_{\xi}U^{\dagger}\partial_{k_x}\mathcal{H}_{\eta}U\mathcal{T}_{\xi})_{nm} (x\rightarrow y,m\leftrightarrow n)}{[(\boldsymbol{\Sigma}E_{\xi})_{nn}-(\boldsymbol{\Sigma}E_{\xi})_{mm}]^2}\nonumber\\
=&2\mathrm{Im}\sum^{4}_{\substack{m=1\\ m\neq n}}\frac{ (\boldsymbol{\Sigma}\mathcal{T}_{\eta}^{\dagger}\partial_{k_x}\mathcal{H}_{\eta}\mathcal{T}_{\eta})_{nm} (x\rightarrow y,m\leftrightarrow n)}{[(\boldsymbol{\Sigma}E_{\eta})_{nn}-(\boldsymbol{\Sigma}E_{\eta})_{mm}]^2}\nonumber\\
=&-\Omega_{n\eta}.
\end{align}

By summing the Berry curvature which is antisymmetric with respect to the $x$ axis over the FBZ, it can be readily found from Eq.~\eqref{EQ:Chern number} and Eq.~\eqref{EQ:KXY} that $\mathcal{C}_n = 0$ and $\kappa_{ab} = 0$.

\begin{figure}[!ht]
    \centering
    \includegraphics[width=0.95\linewidth]{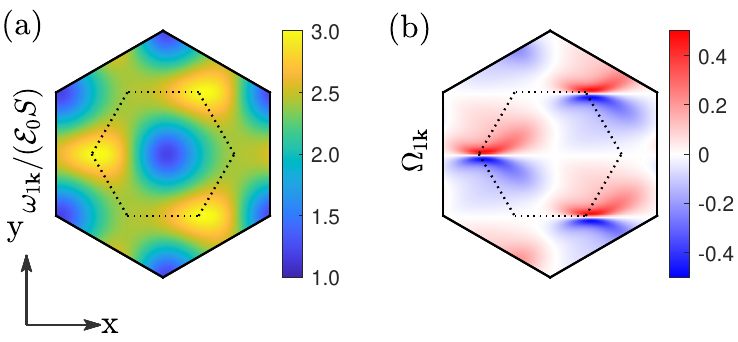}\\
    \caption{(a) The lower magnon band (a) dispersion and (b) Berry curvature at the parameter point $(\psi/\pi, h/(\mathcal{E}_0S)) = (0.1,1.4)$ in the $b$-directional magnetic field.
    }\label{Fig:AFM Energy_and Cur}
\end{figure}

Additionally, we numerically verify the Theorem 1 by plotting magnon dispersion and Berry curvature at the parameter point $(\psi/\pi, h/(\mathcal{E}_0S)) = (0.1,1.4)$ in the $b$-directional magnetic field.
As can be seen from Fig.~\ref{Fig:AFM Energy_and Cur}, the magnon dispersion and Berry curvature are symmetric and antisymmetric with respect to the $x$ axis, respectively.
This is consistent with the conditions that $E_{\xi} = E_{\eta}$ and $\Omega_{n\xi} = -\Omega_{n\eta}$.
\subsection{Topological phase transitions in general directions}\label{chapter3sectionC}
In the $JK\Gamma\Gamma'$ model [cf. Eq.~\eqref{JKGGpHc-Ham} without magnetic field], its symmetric group is $G = D_{3d} \times \mathbb{Z}_2^{\mathrm{T}}$ \cite{Wang2019PRL}, which includes a time-reversal symmetry, a $C_3$ symmetry about the $c$ axis, and a $C_2$ symmetry about the $b$ axis \cite{Nandini2021PRB}.
In the AFM spin-flop phase,
(i) by the action of time reversal, i.e., $\varphi\rightarrow\varphi+\pi$, the spins change as $\mathbf{S}_i \rightarrow -\mathbf{S}_i$ and $\mathcal{H}_{\bf k}$ in Eq.~\eqref{EQ:BdGHam} changes as $(\mathcal{H}_{-\bf k})^* \rightarrow \mathcal{H}_{\bf k}$, the Chern number changes sign \cite{Zhang2021PRB}.
(ii) By $C_3$ rotation of the field $\mathbf{h}$, i.e., $\varphi\rightarrow\varphi+2\pi/3$, the Chern number remains invariant.
(iii) By $C_2$ rotation of the field $\mathbf{h}$ about $b$ axis, i.e., $\varphi\rightarrow\pi-\varphi$, the spins change as $\mathbf{S}_i(\theta,\phi) \rightarrow \mathbf{S}_i(\pi-\theta,\pi-\phi)$ and the momentum coordinates in $\mathcal{H}_{\bf k}$ change as $k_x \rightarrow -k_x$, the Chern number changes sign \cite{Chern2021PRL,Chern2024PRB}.
Therefore, it is sufficient to study field angles in the range $0 \leq \varphi \leq \pi/6$, as all other angles can be derived through symmetries.
The topological phase diagrams of the AFM spin-flop phase at field angles $\varphi$ = $\pi/24$, $\pi/12$, and $\pi/8$, and an animation showing the $\varphi$-dependence with one-degree increments, are provided in Sec.~S4 of the Supplemental Material \cite{SuppMat}.
\section{FM spin-flop phase}\label{chapter4}
\subsection{Energy level splitting and band-gap closing}\label{chapter4sectionA}
\begin{figure}[htbp]
    \centering
    \includegraphics[width=0.95\linewidth]{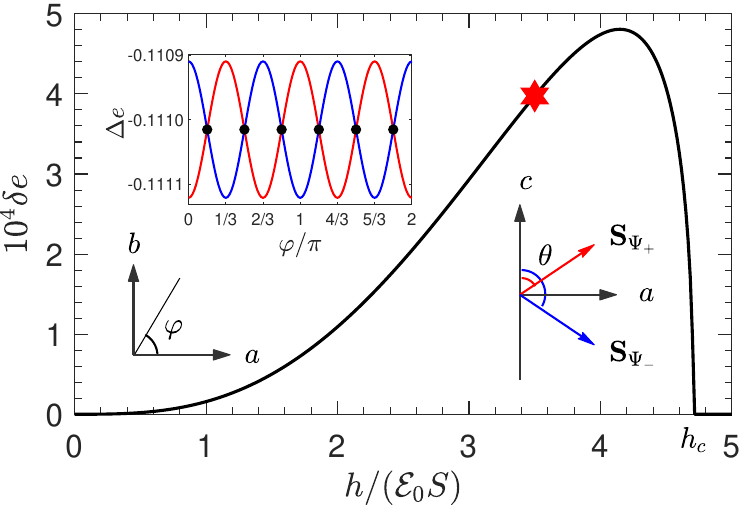}
    \caption{The energy barrier $\delta e$ as a function of the magnetic field $h$ at the parameter point $\psi/\pi=1.1$ in the $a$-directional magnetic field.
    Inset: Spin-wave energy correction $\Delta e$ of the $\Psi_+$ state (red) and $\Psi_-$ state (blue) at the parameter point $(\psi/\pi,h/(\mathcal{E}_0S))=(1.1, 3.5)$ vs field angle $\varphi$. The black dots represent the directions of $b$ and its equivalent ones.}
    \label{Fig:degenerate}
\end{figure}

In contrast to the AFM spin-flop phase, the degeneracy of the two classically equivalent FM spin-flop phases is slightly lifted due to quantum fluctuations.
For concreteness, we denote the two states as $\Psi_+$ and $\Psi_-$, respectively, and the spins are $S(\theta, \phi)$ and $S(\pi-\theta, \phi)$ with $\theta \in (0, \pi/2)$.
To discern the energy splitting, we calculate the energy reduction $\Delta e(\varphi) = e_{\mathrm{sw}}(\varphi) - S^2e_{\mathrm{cl}}$ with respect to the in-plane angle $\varphi$ and a typical example with $\psi/\pi = 1.1$ and $h/(\mathcal{E}_0S) = 3.5$ is shown in the inset of Fig.~\ref{Fig:degenerate}.
It is observed that $\Delta e(\varphi)$ in each state has a period of $2\pi/3$ and its extrema are equal to each other..
The difference lies in that their values are in general different except for six accidental points at $\varphi/\pi = 1/6 + n/3$ $(n = 0\to5)$, indicating that one state is more energetically favored over the other.
Specifically, the energy barrier defined as $\delta e = \lvert\Delta e^{\Psi_+}(\varphi) - \Delta e^{\Psi_-}(\varphi)\rvert$, reaches its maximum value at $\varphi/\pi = n/3$ $(n = 0\to5)$.
\begin{table*}
\centering
\caption{Summary of all band-gap closing scenarios for the FM spin-flop phase under an in-plane magnetic field. It includes the band-gap closing points, the corresponding colors drawn in Fig.~\ref{Fig:FM diagram}, and the constraint equations for the band-gap closing.}
\begin{ruledtabular}
\begin{tabular}{c c c c}
Lines&Points& Colors & Constraint equations \\
\colrule
\rule{0pt}{1.2em}
\!\!1&$\boldsymbol{\Gamma}$ & \begin{tikzpicture}
\draw (0,0)--(1.2,0)[color=clr_Gamma][line width=4pt];
\end{tikzpicture}
& $3\sin^2\theta-2=0$ \\
2&$\textbf{M}_1$ & \begin{tikzpicture}
\draw (0,0)--(1.2,0)[color=clr_M1][line width=4pt];
\end{tikzpicture}
& $\widetilde\Gamma(3\sin^2\theta-2)-4\overline\Gamma\big[\sin^2\theta\cos(2\varphi-\pi/3)+\sqrt{2}\sin\theta\cos\theta\cos(\varphi+\pi/3)\big]=0$ \\
3&$\textbf{M}_2$ & \begin{tikzpicture}
\draw (0,0)--(1.2,0)[color=clr_M2][line width=4pt];
\end{tikzpicture}
& $\widetilde\Gamma(3\sin^2\theta-2)+4\overline\Gamma(\sin^2\theta\cos2\varphi+\sqrt{2}\sin\theta\cos\theta\cos\varphi)=0$
\\
4&$\textbf{M}_3$ & \begin{tikzpicture}
\draw (0,0)--(1.2,0)[color=clr_M3][line width=4pt];
\end{tikzpicture}
& $\widetilde\Gamma(3\sin^2\theta-2)-4\overline\Gamma\big[\sin^2\theta\cos(2\varphi+\pi/3)+\sqrt{2}\sin\theta\cos\theta\cos(\varphi-\pi/3)\big]=0$
\\
5&$\overline{\textbf{M}_2 \textbf{K}'},\overline{\textbf{KV}'}$ & \begin{tikzpicture}
\draw (0,0)--(1.2,0)[color=clr_FMS1][line width=4pt];
\end{tikzpicture}
& $\sqrt{2}\tan\theta\cos\widetilde{\varphi}=1$, where $\widetilde{\varphi}=\mathrm{mod}(\varphi+\pi/6,\pi/3)-\pi/6$
\end{tabular}
\end{ruledtabular}
\label{Tab:FMSpinFlop}
\end{table*}
Therefore, when the magnetic field is aligned along the $b$ or its equivalent directions, the two states remain degenerate in the context of LSWT as the spin-wave energies are equal due to a $C_2$ symmetry.
In contrast, when the magnetic field is aligned along the $a$ or its equivalent directions, the energy barrier gets its maximal value.
The main panel of Fig.~\ref{Fig:degenerate} shows the energy barrier $\delta e$ as a function of $h/(\mathcal{E}_0S)$ for $\psi/\pi = 1.1$ in the $a$-directional magnetic field.
It gradually increases in the FM spin-flop phase, drops rapidly when approaching the PM, and eventually vanishes within the PM.
We also calculate the energy barrier $\delta e$ at other exchange parameter $\psi$ and magnetic field $h$, and find that the lower energy state depends solely on the field angle $\varphi$.

First, we study the band-gap closing of the FM spin-flop phase.
Similarly, we introduce eight real numbers incorporating spin angle $\theta \in (0,\pi)$ and field angle $\varphi \in [0,2\pi)$, which are
$c_1=\widetilde\Gamma(3\sin^2\theta-2)+\overline\Gamma(\sin^2\theta\cos2\varphi+\sin2\theta\cos\varphi/\sqrt{2})$,
$c_2=3\widetilde\Gamma(3\sin^2\theta-2)-3c_1$,
$c_3=\sqrt{3}\,\overline\Gamma(\sin^2\theta\sin2\varphi-\sin2\theta\sin\varphi/\sqrt{2})$,
$c_4=3\widetilde\Gamma\sin^2\theta+\overline\Gamma\big[(\sin^2\theta-2)\cos2\varphi+\sin2\theta\cos\varphi/\sqrt{2}\big]$,
$c_5=9\widetilde\Gamma\sin^2\theta-3c_4$,
$c_6=\sqrt{3}\,\overline\Gamma\big[(\sin^2\theta-2)\sin2\varphi-\sin2\theta\sin\varphi/\sqrt{2}\,\big]$,
$c_7=\overline\Gamma(\cos\theta\sin2\varphi-\sin\theta\sin\varphi/\sqrt{2})$, and
$c_8=-\sqrt{3}\,\overline\Gamma(\cos\theta\cos2\varphi+\sin\theta\cos\varphi/\sqrt{2})$.
Thus, $\lambda_0(\mathbf{k})$ and $\lambda_1(\mathbf{k})$ in Table~\ref{Tab-LSWTBdG} are simplified to
\begin{subequations}
\begin{align}
\!\!\!\lambda_0(\mathbf{k}) \!&= \!\frac{1}{6}(c_1f_{\mathbf{k}}^*+c_2+c_3g_{\mathbf{k}}^*), \label{EQ:FMFldALmabda0} \\
\!\!\!\lambda_1(\mathbf{k}) \!&= \!\frac{1}{6}\Big\{\!c_4f_{\mathbf{k}}^*\!+\!c_5\!+\!c_6g_{\mathbf{k}}^*\!+\!2i\big[c_7(f_{\mathbf{k}}^*-3)\!+\!c_8g_{\mathbf{k}}^*\big]\!\Big\}. \label{EQ:FMFldALmabda1}
\end{align}
\end{subequations}
When $c_3\neq 0$, $\Delta_2$ in Eq.~\eqref{EQ:Del1Del2} turns to \cite{Chern2024PRB}
\begin{equation}\label{EQ:FMSFDelta2}
\Delta_2 = \left(\frac{C}{9c_3}\right)^2(\sin k_1+\sin k_2)^2,
\end{equation}
where
\begin{equation}\label{EQ:FMSFDelta2Coeff}
C = c_8(c_2c_4-c_1c_5)+c_7\big[c_3(3c_4+c_5)-c_6(3c_1+c_2)\big].
\end{equation}
Akin to the strategy used in the AFM spin-flop phase, the way to identify the band-gap closing is to ensure $\Delta_1 = 0$ and $\Delta_2 = 0$.
In the former, it is inferred that
\begin{subequations}\label{EQ:FMFldALambda0RealImag}
\begin{align}
&c_1(1+\cos k_1+\cos k_2)+c_2=c_3(\cos k_2-\cos k_1), \label{EQ:FMReal}\\
&c_1(\sin k_1+\sin k_2)-c_3(\sin k_1-\sin k_2)=0, \label{EQ:FMImag}
\end{align}
\end{subequations}
while in the latter, one should have $\sin k_1+\sin k_2=0$ or $C = 0$.
Therefore, this leads to two categories of solutions.
In the first category, $(k_1,k_2) \in$ $\big\{\boldsymbol{\Gamma}(0,0),\textbf{M}_1(\pi,0),\\\textbf{M}_2(\pi,\pi),\textbf{M}_3(0,\pi)\big\}$ by $\sin k_1=\sin k_2=0$. Eq.~\eqref{EQ:FMImag} and Eq.~\eqref{EQ:FMSFDelta2} is satisfied, and Eq.~\eqref{EQ:FMReal} leads to the constraint equations $3c_1+c_2 = 0$, $c_1+c_2-2c_3 = 0$, $c_1-c_2 = 0$, and $c_1+c_2+2c_3 = 0$, respectively.
By contrast, in the second category where $(k_1,k_2) \notin$ $\{\boldsymbol{\Gamma},\textbf{M}_1,\textbf{M}_2,\textbf{M}_3\}$, by satisfying $C=0$ or $c_3=0$, the resultant constraint equation on the angles of spin $\theta$ and field $\varphi$ is
\begin{equation}
\sqrt{2}\tan\theta\cos\widetilde{\varphi} = 1
\end{equation}
where $\widetilde{\varphi}=\mathrm{mod}(\varphi+\pi/6,\pi/3)-\pi/6$.
Hence, the value of $\theta$ is obtained by solving the constraint equation when $\psi/\pi \geq 1.1476$ (i.e., $\Gamma > 2\Gamma'$),
and the topological phase boundary is further determined via Eq.~\eqref{EQ:PolarAngle}.
The loci of the band-gap closing points in the reciprocal space and the constraint equations for the band-gap closing are summarized in Table~\ref{Tab:FMSpinFlop}.

To proceed further, we study the topological phase transitions in the FM spin-flop phase when the magnetic field is along the $a$, $b$, and general directions in sequence.
\subsection{Topological phase transitions when $\mathbf{h}\,||\,a [11\bar2]$}\label{Sec4:ADir}
\begin{figure*}[htbp]
    \centering
    \includegraphics[width=0.95\linewidth]{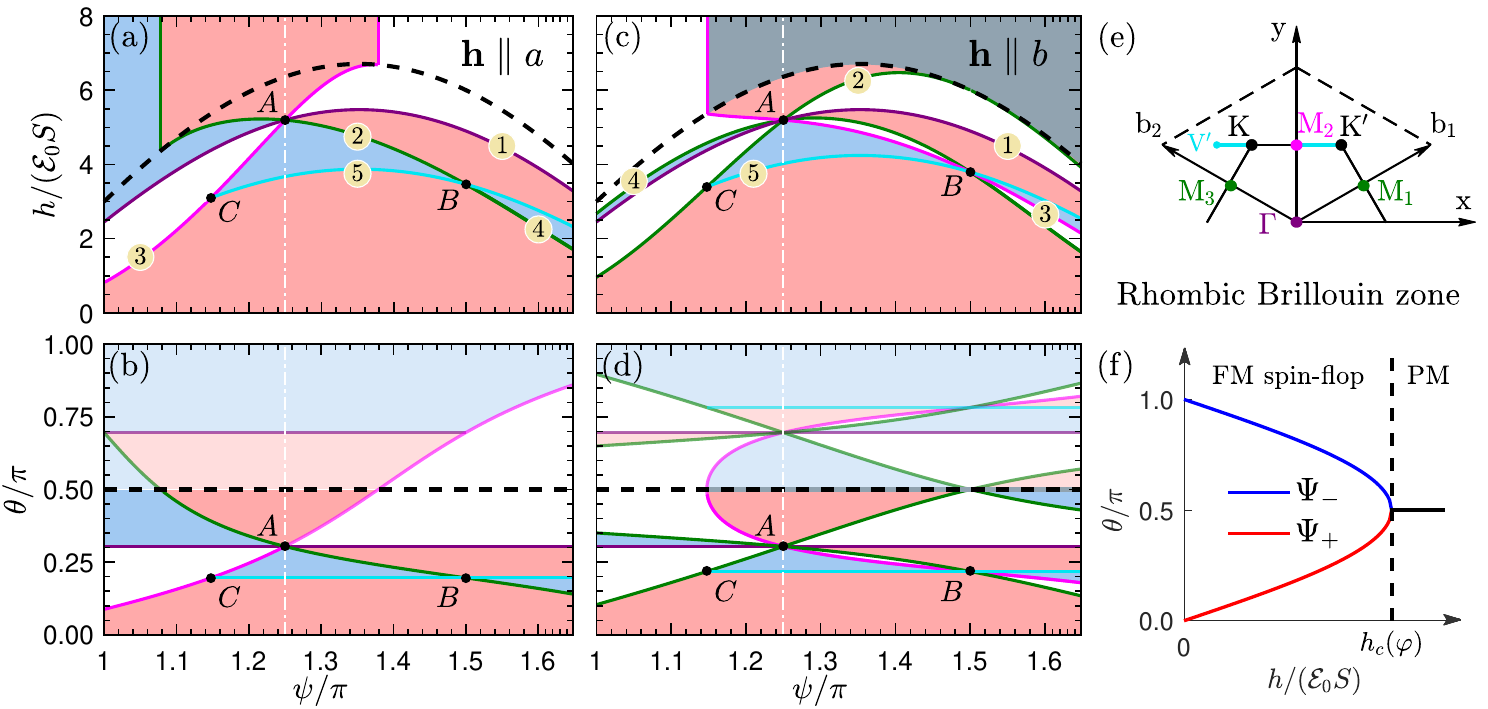}
    \caption{Topological phase diagrams of (a) the $\Psi_+$ state and (b) both the $\Psi_+$/$\Psi_-$ states in the FM spin-flop phase in the $a$-directional magnetic field.
    The letters ($A$, $B$, and $C$) denote the confluence points while the digits from 1 to 5 mark different topological phase transition lines.
    The $y$ labels in panels (a) and (b) are the magnetic field $h$ and the spin polar angle $\theta$, respectively.
    The black dashed line at $\theta/\pi=0.5$ in panel (b) is the dividing line between the $\Psi_+$ and $\Psi_-$ states. The band-gap closing points and paths are labeled in (e) with the same color.
    (c-d) The similar topological phase diagrams as these in (a-b) but for the $b$-directional magnetic field.
    The gray area in panel (c) indicates the Chern number is ill-defined due to the closure of band gap.
    (e) The rhombic Brillouin zone with the relevant high-symmetry points.
    (f) Schematic diagram of the spin polar angle $\theta$ and the magnetic field $h$ in $\Psi_+$ (red) and $\Psi_-$ (blue) states.}
    \label{Fig:FM diagram}
\end{figure*}

When the magnetic field is aligned along the $a$ direction, the spin-wave energy $e_{\mathrm{sw}}$ in $\Psi_+$ state is lower.
Figure~\ref{Fig:FM diagram}(a) shows the topological phase diagram of the $\Psi_+$ state.
When $h = 0$, the ground state is the FM$_c$ phase whose spins point along the $c$ direction.
The band-gap opens, and the Chern number is found to be $\mathcal{C}_1 = +1$ \cite{McClarty2018PRB}.
When $h > h_c$, the ground state is divided into three parts whose Chern numbers are
$\mathcal{C}_1 = -1$ ($1.0 < \psi/\pi < 1.078$), $\mathcal{C}_1 = +1$ ($1.078 < \psi/\pi < 1.379$), and $\mathcal{C}_1 = 0$ ($1.379 < \psi/\pi < 1.648$), which is consistent with the parameter region of the AFM spin-flop phase \cite{Chern2024PRB}.
When $0<h < h_c$, the topological phase transitions in the FM spin-flop phase are much richer compared to the AFM spin-flop phase.
According to Table~\ref{Tab:FMSpinFlop}, the band-gap closing points and the associated topological transition lines can be classified into five categories.
They intersect at points $A (1.25, 3\sqrt3)$ and $B (1.50, 2\sqrt3)$ in the parameter spaces of $(\psi/\pi, h/(\mathcal{E}_0S))$.
Of note is that the line 5, whose band-gap closing point is illustrated in Fig.~\ref{Fig:FM diagram}(e), terminates at point $C (1.1476, 4\sqrt{3/5})$ in connection to line 3.
To further unravel the topological transition lines, we present the full topological phase diagram of both $\Psi_+$ and $\Psi_-$ states denoted by spin $\theta$ in Fig.~\ref{Fig:FM diagram}(b) and the relationship between the $\theta$ and $h$ can be found in Fig.~\ref{Fig:FM diagram}(f) and Eq.~\eqref{EQ:PolarAngle}.
We find that the spin angle $\theta$ of lines 1 and 5 are fixed at $\tan^{-1}(\sqrt{2})\approx 0.3041\pi$ and $\tan^{-1}(1/\sqrt{2})\approx 0.1959\pi$ in $\Psi_+$ state, respectively.
There are only three topological transition lines in $\Psi_-$ state.
One is fixed at $\theta = \pi-\tan^{-1}(\sqrt{2}) \approx 0.6959\pi$ (which is symmetric to line 1 with respect to the middle), while the others are extended lines of line 2 and line 3.

\begin{figure}[!ht]
    \centering
    \includegraphics[width=0.95\columnwidth]{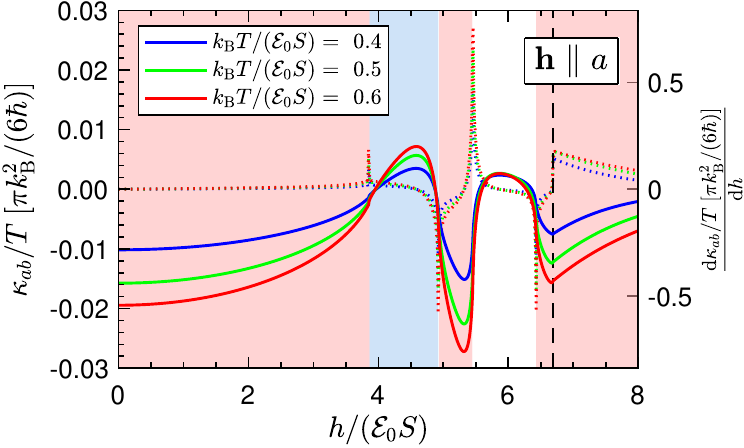}
    \caption{Thermal Hall conductivity $\kappa_{ab}/T$ [in units of $\pi k^2_\mathrm{B}/(6\hbar)$] (solid curves) and its first derivative $\mathrm{d}(\kappa_{ab}/T)/\mathrm{d} h$ (dotted curves) vs magnetic field $h$ in the parameter of $\psi/\pi = 1.325$.
    The magnetic field is along the $a$ direction and spin $S = 1/2$ is used.
    The red, white, and blue areas indicate $\mathcal{C}_1$ = $+1$, $0$, and $-1$, respectively.
    The black dashed curve denotes the magnetic phase transition to the PM.}
    \label{Fig:FM_kxy}
\end{figure}
Figure \ref{Fig:FM_kxy} shows the thermal Hall conductivity $\kappa_{ab}$ and its first derivative as a function of field strength $h/(\mathcal{E}_0S)$ for $\psi/\pi = 1.325$, which encompass many topological phase transitions of different kinds.
To begin with, the sign of $\kappa_{ab}$ is negative in the PM since the underlying Chern number $\mathcal{C}_1$ equals to $+1$.
Similar to the AFM spin-flop phase, $\kappa_{ab}/T$ exhibits a kink and its derivative has a jump at the magnetic phase transition.
By decreasing the magnetic field, the Chern number $\mathcal{C}_1$ changes from positive to zero.
In the topologically trivial region where $5.46 < h/(\mathcal{E}_0S) < 6.42$, it is observed that the Berry curvature in the lower part of band is primarily negative.
Thus, the thermal Hall conductivity $\kappa_{ab}$ enjoys a sign change near the topological phase transitions.
As the magnetic field further decreases, $\kappa_{ab}$ undergoes several sign changes, signifying a couple of topological phase transitions.
Furthermore, there is a sharp peak in the derivative of $\kappa_{ab}$ when the magnetic field is close to the topological phase transitions.

On the other hand, appearance of chiral edge modes due to nontrivial band topology is another hallmark of topological phases.
When the open boundary condition is adopted in one of the two directions in the honeycomb lattice, there will be likely chiral edge modes concatenating the upper and lower bands \cite{Mook2014PRB}.
According to the bulk-edge correspondence \cite{Yasuhiro1993PRL}, since the FM spin-flop phase only has two magnon bands, the number of pairs of edge modes in the intermediate band-gap equals the value of the Chern number $\mathcal{C}_1$, and their propagation direction is determined by the sign of the Chern number $\mathrm{sgn}(\mathcal{C}_1)$.
\begin{figure}[htbp]
    \centering
    \includegraphics[width=0.95\columnwidth]{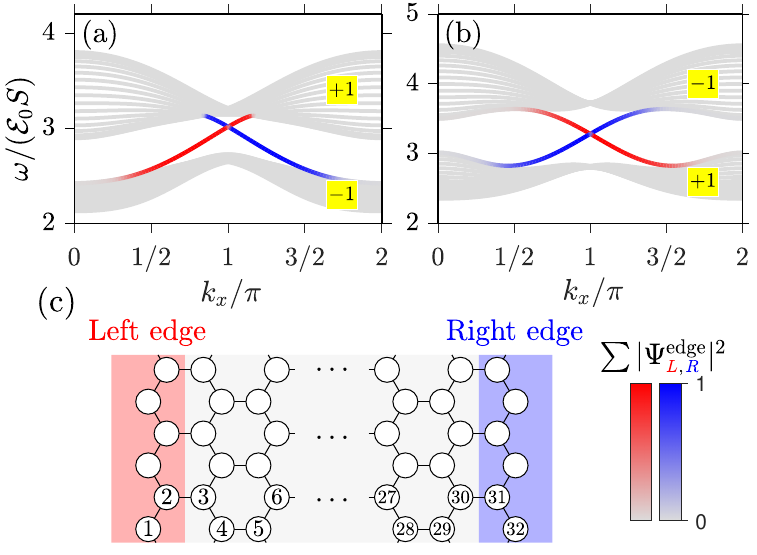}
    \caption{The magnon bands in the zigzag nanoribbon geometry at (a) $\psi/\pi = 1.10$ and (b) $\psi/\pi = 1.55$ in the $a$-directional magnetic field with $h/(\mathcal{E}_0S) = 4.2$. Probability density of the left (red) and right (blue) edge [defined as the two outermost layers at each edge side, see panel (c)] is encoded by color. (c) Section of the zigzag nanoribbon geometry with 32 layers including layer labels, left (red), and right (blue) edge.
    The system is applied with open (periodic) boundary condition along the armchair (zigzag) direction.
    }\label{fig:edge}
\end{figure}

Figure~\ref{fig:edge} shows the magnon band in the zigzag nanoribbon geometry with (a) $\psi/\pi = 1.10$ and (b) $\psi/\pi = 1.55$, which are located in the regimes of $\mathcal{C}_1$ = $-1$ and $+1$, respectively (The computational procedure is shown in Sec.~S5 in the Supplemental Material \cite{SuppMat}).
Because $\vert\mathcal{C}_1\vert = 1$ for both cases, it is observed that there are only one pair of edge modes in the bulk band gap severally.
In addition, the intensity of the curve color denotes the probability density of the wave functions in the left and right edges \cite{Diaz2020PRR}.
The bright color in the edge modes acts as an evidence of the localization of modes, indicating that these edge modes are indeed chiral edge modes.
We observe that, at the same $k_x$ point, the change in the sign of the Chern number $\mathcal{C}_1$ leads to a change in the localization direction of the edge modes.
It also leads to a change in the overall propagation direction of the edge modes.
Noteworthily, we also calculate the dynamic spin structure factor $\mathcal{S}(\mathbf{k},\omega)$ of the FM spin-flop phase in the zigzag nanoribbon geometry \cite{Joshi2018PRB,Janssen2019JPCM}.
The occurrence of the edge modes is rather evident since they exhibit superior intensities (see Sec.~S6 in the Supplemental Material \cite{SuppMat}).
Therefore, it is interesting to observe the chiral edge modes directly in future neutron scattering experiments.

\subsection{Topological phase transitions when $\mathbf{h}\,||\,b [\bar110]$}\label{Sec4:BDir}

When the magnetic field is aligned along the $b$ direction, the spin-wave energies of $\Psi_+$ and $\Psi_-$ are degenerate, Chern numbers $\mathcal{C}_n$ satisfy the following relationship.

\textbf{Theorem 2.} The Chern numbers $\mathcal{C}_n$ and thermal Hall conductivity $\kappa_{ab}$ of the $\Psi_+$ and $\Psi_-$ states in the FM spin-flop phase with the magnetic field along the $b$ direction differ by a minus sign.

\textit{Proof}. We start by defining two symmetric points about the $y$ axis in the reciprocal space, which are $\mathbf{k}(\Psi_+) = (k_x, k_y)$ point in the $\Psi_+$ state and $\mathbf{k}(\Psi_-) = (-k_x,k_y)$ point in the $\Psi_-$ state.
When the field is applied along the $b$ direction, the Hamiltonian possesses a $C_2$ rotational symmetry about the $b$ axis.
Under this operation, the states transform as $\Psi_+\rightarrow\Psi_-$ and the reciprocal space coordinates transform as $k_x\rightarrow-k_x$ in the reciprocal space, while $\mathcal{H}_\mathbf{k}$ remains invariant. Thus,
\begin{align}\label{EQ:FM_H}
\mathcal{H}_{(k_x,k_y)}^{\Psi_+}=\mathcal{H}_{(-k_x,k_y)}^{\Psi_-},
\end{align}
and the corresponding dispersions satisfy
\begin{align}\label{EQ:FM_Epsilon}
\omega^{\Psi_+}_{n(k_x,k_y)} = \omega^{\Psi_-}_{n(-k_x,k_y)}.
\end{align}
Differentiating $\mathcal{H}$ in Eq.~\eqref{EQ:FM_H} with respect to $k_x$ and $k_y$, we can obtain
\begin{align}\label{EQ:FM_H_derivative}
\!\frac{\partial \mathcal{H}^{\Psi_+}_{(k_x,k_y)}}{\partial{k_x}}\!=\!-\frac{\partial \mathcal{H}^{\Psi_-}_{(-k_x,k_y)}}{\partial{k_x}},\!\frac{\partial \mathcal{H}^{\Psi_+}_{(k_x,k_y)}}{\partial{k_y}}\!=\!\frac{\partial \mathcal{H}^{\Psi_-}_{(-k_x,k_y)}}{\partial{k_y}}.
\end{align}
With Eq.~\eqref{EQ:Berry Curvature} in mind, the Berry curvatures of the two states are related by
\begin{align}\label{EQ:FM_Omega}
\Omega^{\Psi_+}_{n(k_x,k_y)}=-\Omega^{\Psi_-}_{n(-k_x,k_y)}.
\end{align}
By summing the Berry curvature which is antisymmetric with respect to the $y$ axis over the FBZ,
we deduce from Eq.~\eqref{EQ:Chern number} and Eq.~\eqref{EQ:KXY} that
$\mathcal{C}^{\Psi_+}_n + \mathcal{C}^{\Psi_-}_n = 0$ and $\kappa_{ab}^{\Psi_+} + \kappa_{ab}^{\Psi_-} = 0$.
This completes the proof of Theorem 2.
\begin{figure}[htbp]
    \centering
    \includegraphics[width=0.95\linewidth]{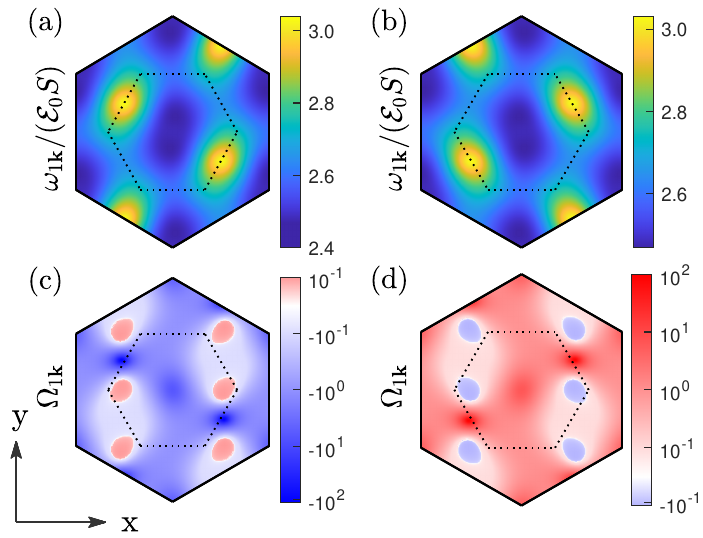}
    \caption{The lower magnon band dispersions and Berry curvatures of (a, c) $\Psi_+$ and (b, d) $\Psi_-$ at the parameter point $(\psi/\pi, h/(\mathcal{E}_0S))=(1.1, 4)$ in the $b$-directional magnetic field.}
    \label{Fig:FMflopBerry}
\end{figure}

Therefore, when the magnetic field is aligned along the $b$ direction, the magnon dispersions $\omega_{n\textbf{k}}$ and the Berry curvatures $\Omega_{n\textbf{k}}$ of the $\Psi_+$ and $\Psi_-$ states satisfy particular relations revealed in Eqs.~\eqref{EQ:FM_Epsilon} and \eqref{EQ:FM_Omega}.
As shown in Fig.~\ref{Fig:FMflopBerry}, these relations are numerically verified at the specific parameter $(\psi/\pi, h/(\mathcal{E}_0S))=(1.1, 4)$.
Consequently, irrelevant of the intensity of magnetic field, the Chern numbers and the thermal Hall conductivity at each temperature differ by a sign.

\begin{figure*}[htbp]
    \centering
    \includegraphics[width=0.95\linewidth]{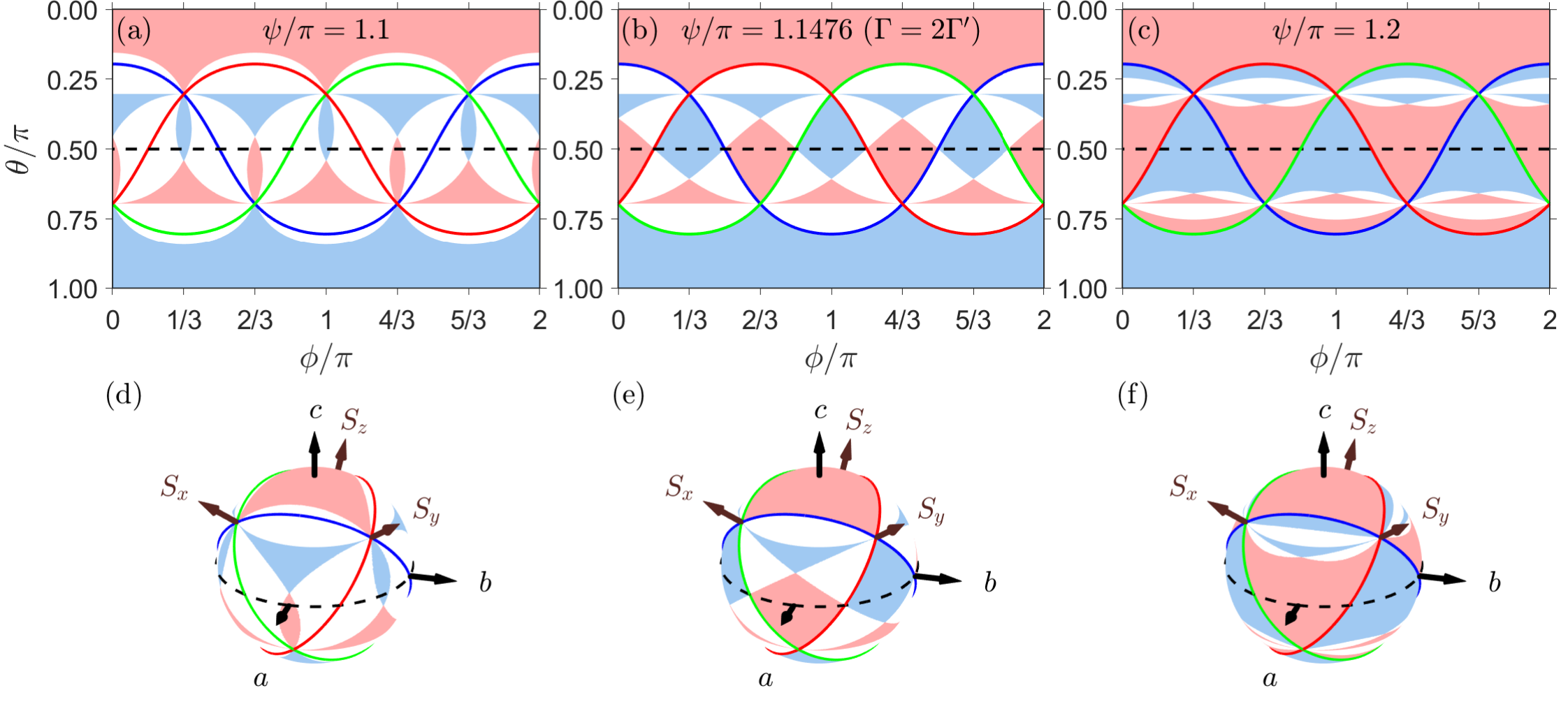}
    \caption{Topological phase diagrams of the FM spin-flop phase on the plane of the spin angles $(\phi, \theta)$ with the exchange coupling angles (a) $\psi/\pi = 1.1$, (b) $\psi/\pi = 1.1476 (\Gamma = 2\Gamma')$, and (c) $\psi/\pi = 1.2$.
    The red, green, and blue curves indicate $S_y\mbox{-}S_z$, $S_x\mbox{-}S_z$, and $S_x\mbox{-}S_y$ spin coordinate planes, respectively. The in-plane magnetic field direction aligns with the spin azimuthal angle, i.e., $\varphi = \phi$.
    (d-f) Spherical plots of topological phase diagrams at the same parameters of these in panels (a-c) correspondingly.
    The upper and lower hemispheres represent $\Psi_+$ and $\Psi_-$ states, respectively.
    The spin coordinate axis $(S_x, S_y, S_z)$ and the crystallographic axis $(a, b, c)$ are presented simultaneously.
    }\label{Fig:FMflopdirall}
\end{figure*}

In contrast to the AFM spin-flop phase, a nontrivial observation is that the Chern number in the FM spin-flop phase is not necessarily zero when $h\!\parallel\!b$.
Figure~\ref{Fig:FM diagram}(c) shows the topological phase diagram for the $\Psi_+$ state of the FM spin-flop phase.
It is found that the landscape resembles that of the panel (a) in which the magnetic field is along the $a$ direction.
The five topological transition lines are marked accordingly, and the coordinates for the three confluence points are
$A (1.25, 3\sqrt3)$, $B (1.50, 6\sqrt{2/5})$, and $C (1.1476, 12\sqrt{2}/5)$, in the parameter space of $(\psi/\pi, h/(\mathcal{E}_0S))$.
As the magnetic field direction changes, the general expressions for these three points are $A (1.25, 3\sqrt3)$, $B (1.50, 6/\sqrt{2\cos^2\widetilde{\varphi} + 1})$, and $C (1.1476, 12/\sqrt{5(2\cos^2\widetilde{\varphi} + 1)})$.
Also of note is that the Chern number in the PM is either zero or ill-defined.
Figure~\ref{Fig:FM diagram}(d) shows the topological phase diagram as a function of $\theta$, which depicts both the $\Psi_+$ and $\Psi_-$ states.
In accordance with Theorem 2, it is observed that the topological phase diagram is symmetric in the sense of $\theta \to \pi-\theta$ and $\mathcal{C}_1 \to -\mathcal{C}_1$.

\subsection{Topological phase transitions in general directions}\label{Sec4:Generic}
Finally, we discuss the topological phase transitions in the parameter space defined by the in-plane field angle $\varphi$ and polar angle of spin $\theta$,with a fixed exchange coupling angle $\psi$.
Figure~\ref{Fig:FMflopdirall} shows the topological phase diagrams at the exchange coupling angles $\psi/\pi = 1.1, 1.1476 (\Gamma = 2\Gamma')$, and $1.2$, and the $S_y\mbox{-}S_z$ (in red), $S_x\mbox{-}S_z$ (in green), and $S_x\mbox{-}S_y$ (in blue) spin coordinate planes are shown for comparison.
In panels (a) and (d) where $\psi/\pi = 1.1$, several segments of these coordinate axes planes are close to, but not situated at the topological phase transition boundaries.
As the exchange coupling angle $\psi$ increases, some topological phase transitions coincide with the spin coordinate planes when $\psi/\pi \geq 1.1476$ (see other panels).
This implies that, under the influence of an in-plane magnetic field, a topological phase transition occurs when the spin crosses one of the spin coordinate planes.

Additionally, the topological phase diagrams of the FM spin-flop phase at field angles $\varphi$ = $\pi/24$, $\pi/12$, and $\pi/8$, and an animation showing the $\varphi$-dependence with one-degree increments, are provided in Sec.~S4 of the Supplemental Material \cite{SuppMat}.
\section{Conclusion and Remark}\label{chapter5}
In summary, we have investigated topological magnons of the coplanar AFM spin-flop phase and collinear FM spin-flop phase in the $\Gamma$-$\Gamma'$ model under the in-plane magnetic field within the framework of LSWT.
By calculating the Chern number, we map out various topological phase diagrams in the parameter space of exchange coupling angle $\psi$ and magnetic field $h$ at different in-plane field angles $\varphi$, with a focus on the $a$ direction ($\varphi = 0$) and $b$ direction ($\varphi = \pi/2$).
After a careful inspection of the band-gap closing conditions, we manage to fathom out the topological phase boundaries and the associated band-gap closing points.
We propose that the thermal Hall conductivity can serve as a tool to perceive different topological phase transitions.
Finally, we use the FM spin-flop phase as an example to investigate features of chiral edge states in two different topological phases, with Chern numbers of the two magnon bands being $(-1, +1)$ and $(+1, -1)$ from bottom to top, respectively.

In the AFM spin-flop phase, three topological phase transition lines stemming from the hidden $U(1)$ symmetric point at $\psi/\pi = 0.25$ when $\mathbf{h}\,||\,a$.
One of the band-gap closing points is located at $\textbf{M}$ point, while the remaining are special paths in the Brillouin zone.
Topological phase transitions can be recognized by the thermal Hall conductivity and its derivative as well.
Specifically, there is a jump, which becomes more evident with the increase of temperature, in the thermal Hall conductivity when crossing the trivial-nontrivial topological phase transition.
For the topological phase transition between two distinct topological phases with different patterns of Chern number, the thermal Hall conductivity is smoothly varied but its derivative exhibits a tendency of divergence.
When $\mathbf{h}\,||\,b$, we prove exactly that the Chern number and thermal Hall conductivity are zero in the whole parameter regions.

In the FM spin-flop phase, LSWT calculation indicate that quantum fluctuations will typically lift the degeneracy of the two classically equivalent configurations.
Hinging on the in-plane field angle, the two magnetic states compete and the ground state equiprobably chooses one of the two with a period of $2\pi/3$.
Nevertheless, the degeneracy remains intact at six special angles $\varphi/\pi = 1/6 + n/3$ ($n = 0, 1, \cdots, 5$).
The Chern numbers of the two degenerate states differ by exactly a minus sign at the same exchange parameters.
In the typical topological phase diagrams, five topological phase transition lines intersect at three confluence points: $A (1.25, 3\sqrt3)$, $B (1.50, 6/\sqrt{2\cos^2\widetilde{\varphi} + 1})$, and $C (1.1476, 12/\sqrt{5(2\cos^2\widetilde{\varphi} + 1)})$.
The band-gap closing points of the five transition lines are the $\boldsymbol{\Gamma}$ point, three distinct $\textbf{M}$ points, and two special segments in reciprocal space.
Clearly, the topological phase transitions in the FM spin-flop phase are much richer than that of the AFM analogue since the former occupy areas triple as large.
We also observe that some topological phase transitions coincide with the coordinate axis planes when the exchange coupling angle $\psi/\pi \geq 1.1476$.

In the future, there are a couple of issues worth further research to advocate the interesting topological magnons in the AFM and FM spin-flop phases.
First, while the LSWT is a reasonable approximation of the low-lying energy spectrum in the two-sublattice phases, the magnon-magnon interaction should be non-negligible when the temperature is elevated, with the possibility of altering band topology and causing physical effects like magnon damping \cite{Smit2020PRB,Choi2023PRB,Koyama2024PRB}.
For another, recent efforts have revealed that the magnon polaron, which comes from the hybridization of magnon and phonon \cite{Sun2023PRB,Kwon2024PRB,Gillig2023PRB}, is able to generate nontrivial band topology and thus has an impact on the thermal Hall measurements.
While being beyond the scope of current work, it is imperative to study the topological aspects of the magnon-magnon interaction and magnon-phonon interaction in the two spin-flop phases.
On all accounts, our findings shed light on the complexity of topological magnons due to bond-dependent interactions and open up a pathway for understanding thermal Hall conductivity in candidate Kitaev materials.

\begin{acknowledgements}
We would like to thank K. Chen for helpful discussions.
We also appreciate the anonymous Referee for pointing out a mistake in deriving the analytical topological phase boundaries.
Q.L. is supported by the National Natural Science Foundation of China (Grants No. 12304176 and No. 12274183)
and the Natural Science Foundation of Jiangsu Province (Grant No. BK20220876).
J.Z. is supported by the National Program on Key Research Project (Grant No. MOST2022YFA1402704)
and the National Natural Science Foundation of China (Grants No. 12274187 and No. 12247101).
X.L. is supported by the Postgraduate Research \& Practice Innovation Program of Jiangsu Province (Grant No. xcxjh20232102).
Q.L. also acknowledges the startup Fund of Nanjing University of Aeronautics and Astronautics (Grant No. YAH21129).
The computations are partially supported by High Performance Computing Platform of Nanjing University of Aeronautics and Astronautics.
\end{acknowledgements}

%




\clearpage

\onecolumngrid

\newpage

\newcounter{sectionSM}
\newcounter{equationSM}
\newcounter{figureSM}
\newcounter{tableSM}
\stepcounter{equationSM}
\setcounter{section}{0}
\setcounter{equation}{0}
\setcounter{figure}{0}
\setcounter{table}{0}
\setcounter{page}{1}
\makeatletter
\renewcommand{\thesection}{\textsc{S}\arabic{section}}
\renewcommand{\theequation}{\textsc{S}\arabic{equation}}
\renewcommand{\thefigure}{\textsc{S}\arabic{figure}}
\renewcommand{\thetable}{\textsc{S}\arabic{table}}


\begin{center}
{\large{\bf Supplemental Material for\\
``Successive topological phase transitions in two distinct spin-flop phases on the honeycomb lattice''}}
\end{center}
\begin{center}
Xudong Li$^{1,\;2}$, Jize Zhao$^{3,\;4}$, Jinbin Li$^{1,\;2}$, and Qiang Luo$^{1,\;2}$ \\
\quad\\
$^1$\textit{College of Physics, Nanjing University of Aeronautics and Astronautics, Nanjing, 211106, China}\\
$^2$\textit{Key Laboratory of Aerospace Information Materials and Physics (NUAA), MIIT, Nanjing, 211106, China}\\
$^3$\textit{School of Physical Science and Technology $\&$ Key Laboratory of Quantum\\ Theory and Applications of MoE, Lanzhou University, Lanzhou 730000, China}\\
$^4$\textit{Lanzhou Center for Theoretical Physics, Key Laboratory of Theoretical\\ Physics of Gansu Province, Lanzhou University, Lanzhou 730000, China}\\
(Dated: January 17, 2025)
\quad\\
\end{center}


\onecolumngrid


\section{Monte Carlo Simulation of $\Gamma$-$\Gamma'$ Model in the In-plane Magnetic Field}\label{SecSM1}
After considering energy-optimized spin configurations of both AFM spin-flop phase and FM spin-flop phase, it is calculated that their classical ground-state energy per site are
\begin{equation}\label{SMEQ:ClEgSpinFlop}
e_{\rm cl}=\left\{
\begin{array}{lcl}
-\widetilde\Gamma-{h^2}/{(2\widetilde\Gamma S^2)},  &      & {\textrm{AFM spin-flop phase}}\\
 \widetilde\Gamma+{h^2}/{(6\widetilde\Gamma S^2)},  &      & {\textrm{FM spin-flop phase}} \\
\end{array} \right.
\end{equation}
where $\widetilde\Gamma = \Gamma + 2\Gamma'$ and $(\Gamma,\Gamma') = \mathcal{E}_0(\cos \psi,\sin \psi)$ ($\mathcal{E}_0 = 1$ is the energy unit).
Also, $\widetilde\Gamma$ is positive~(negative) in AFM~(FM) spin-flop phase.
Similarly, the classical ground-state energy per site in the polarized paramagnet is $e_{\rm cl} = -\widetilde\Gamma/2 - h/S$.

\begin{figure}[!ht]
    \centering
    \includegraphics[width=0.8\columnwidth]{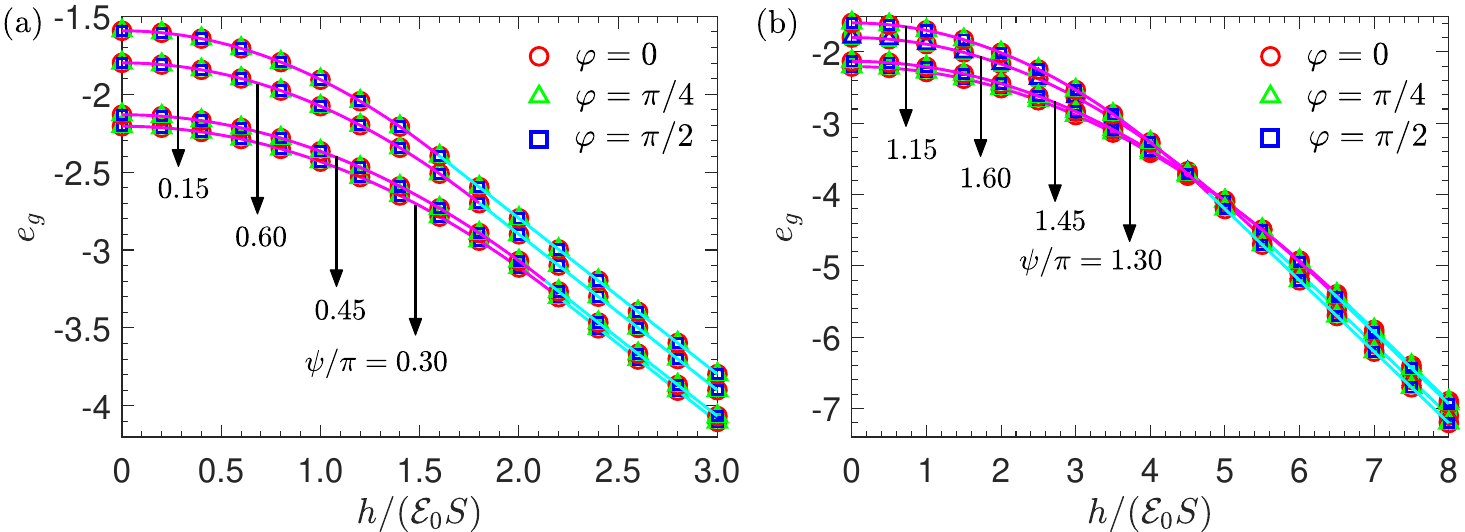}
    \caption{Comparison of the classical energy between the Monte Carlo numerical calculation (symbols) and analytical calculation (curves) in both (a) AFM spin-flop phase and (b) FM spin-flop phase under an in-plane magnetic field.
    The exchange angle $\psi/\pi$ = 0.15, 0.30, 0.45, 0.60 in the AFM spin-flop phase and $\psi/\pi$ = 1.15, 1.30, 1.45, 1.60 in the FM spin-flop phase.
    In each panel, the pink curves represent the energy of spin-flop phase while the cyan curves represent the energy of polarized paramagnet.
    The in-plane field direction is chosen as $a$-direction ($\varphi = 0$, red circle), $b$-direction ($\varphi = \pi/2$, blue square),
    and in between ($\varphi = \pi/4$, green triangle).}
    \label{FigSM:MCEnergy}
\end{figure}

According to Eq.~\eqref{SMEQ:ClEgSpinFlop}, it is inferred that no large-unit-cell ordering appears over the AFM$_c$ and FM$_c$ phases when applying an in-plane magnetic field in the $\Gamma$-$\Gamma'$ model, and the classical energy of the two spin-flop phases does not depend on the in-plane field angle $\varphi$.
To demonstrate this, we thereby perform a series of parallel tempering Monte Carlo simulation on several different clusters including $2\times12\times12$, $2\times16\times16$, $2\times24\times9$, and $2\times24\times15$.
It is found that the ground state is always a sort of spin-flop phase, ruling out the possibility of large-unit-cell ordering.
This is quite different from a previous study by one of the authors in which an out-of-plane field is applied.
In that study, large-unit-cell orderings such as 4-site order and 48-site order are identified at different magnetic field intervals \cite{SMLuo2022PRB}.
In addition, as the in-plane field angle $\varphi$ varies, the ground state does not change significantly except for the direction of spin moment.
Figure~\ref{FigSM:MCEnergy} shows the comparison of the classical energy between the Monte Carlo numerical calculation and analytical calculation in both AFM spin-flop phase and FM spin-flop phase in an in-plane magnetic field.
The fact that analytical results match with Monte Carlo results entirely affirms spin-flop phase as the authentic ground state.

\section{Justification of the Dispersion Relation for AFM spin-flop phase}\label{SecSM2}
\begin{figure}[!ht]
    \centering
    \includegraphics[width=0.8\linewidth]{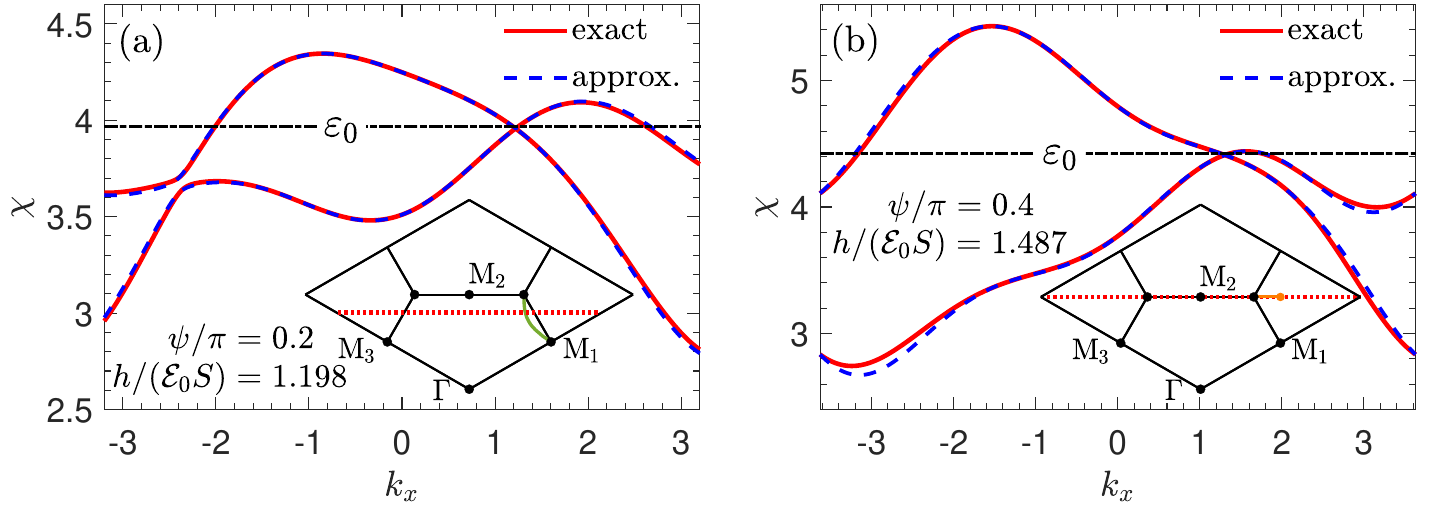}\\
    \caption{Magnon Dispersions $\chi$ near the approximate topological phase transition points: (a) $(\psi/\pi, h/(\mathcal{E}_0S)) = (0.2, 1.198)$, (b) $(\psi/\pi, h/(\mathcal{E}_0S)) = (0.4, 1.487)$.
    The red solid line represents the exact results, the blue dashed line represents the approximate results, and the black dash-dot line represents $\varepsilon_0$.
    The red dotted line in the insets marks the path through the relevant band-gap closing point in momentum space.
    }\label{FigSM:chi}
\end{figure}

In this Section, we analyze the reasonability of approximating the calculation of topological phase boundaries by neglecting the residual term $\mathcal{O} = s(\chi-\varepsilon_0)^2 + t$ in the AFM spin-flop phase.
Expressions for $s$ and $t$ can be found in Eq.~(\textcolor{red}{9}) in the main text.
To begin with, under this approximation, the band-gap closing points satisfy $\Delta_{1} = 4\varepsilon_0^2|\lambda_0(\mathbf{k})|^2 = 0$ $(\varepsilon_0\neq0)$.
This implies $|\lambda_0(\mathbf{k})| = 0$ and also leads to $t = 0$.
Thus, at band-gap closing points, the \textit{second term} of $\mathcal{O}$ disappears.
Figure~\ref{FigSM:chi} illustrates the magnon dispersions $\omega=S\chi/2$ at the topological phase transitions,
The parameters chosen are $(\psi/\pi, h/(\mathcal{E}_0S)) = (0.2, 1.1984)$ and $(\psi/\pi, h/(\mathcal{E}_0S)) = (0.4, 1.4867)$,
which correspond to topological phase transition at line 1 and line 3, respectively.
In each panel, a horizontal path in reciprocal space is chosen to cross the band-gap closing points,
and both the exact and approximate results are shown for comparison.
It can be observed that, at the band-gap closing points, $\chi$ is very close to $\varepsilon_0$.
This indicates that the \textit{first term} of $\mathcal{O}$ is very small.
Taken together, we infer that value of $\mathcal{O}$ is negligibly small at the band-gap closing points.
We therefore conclude that the approximation that neglects the residual term $\mathcal{O}$
can still provide a high-precision result when calculating the topological phase boundaries in the AFM spin-flop phase.

\begin{figure}[!ht]
    \centering
    \includegraphics[width=0.5\linewidth]{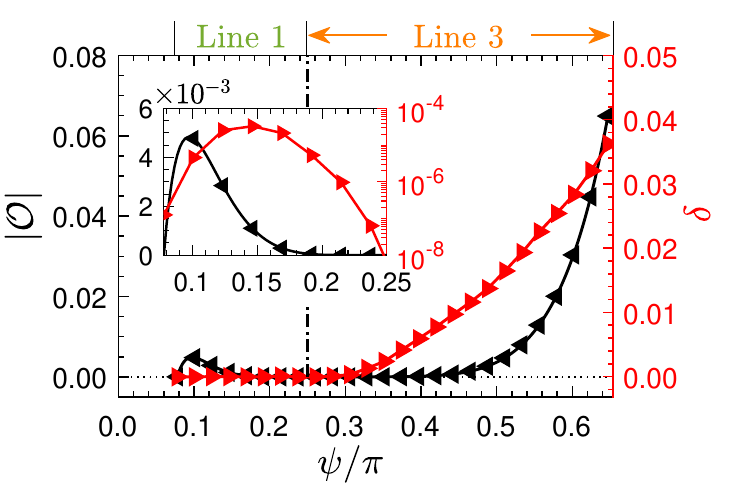}\\
    \caption{The residual term $|\mathcal{O}|$ (in black) and absolute error $\delta$ (in red) in the topological phase boundaries line 1 and line 3 in the AFM spin-flop phase. The topological phase boundaries are indicated at the top. Inset: Zoom in of $|\mathcal{O}|$ and $\delta$ of line 1.
    }\label{Fig:delta}
\end{figure}

Moreover, we also calculate the residual term $|\mathcal{O}|$ and the absolute error $\delta$ between the exact and approximate topological transition points at some exchange angles $\psi$, see Fig.~\ref{Fig:delta}.
We recall that ranges of line 1 and line 3 (cf. Fig.~\textcolor{red}{2} in the main text) are $\psi/\pi \in (0.0780, 0.25)$ and (0.25, 0.6476), respectively.
For the region of line 1, values of both $|\mathcal{O}|$ and $\delta$ are extremely tiny.
For line 2, although $|\mathcal{O}|$ and $\delta$ gradually get bigger as $\varphi$ increases, their values are still reasonably small.
These analysis again suggests that neglecting the residual term $\mathcal{O}$ does not modify the dispersion relations too much.

\section{Proof of $\Delta_3<\Delta_1$ when $\Delta_{1} > 0$ for AFM and FM spin-flop phases}\label{SecSM3}
This Section aims to prove that when $\Delta_{1} = 4\varepsilon_0^2|\lambda_0(\mathbf{k})|^2 > 0$,
the inequality $\Delta_3 < \Delta_1$ always holds in both the AFM spin-flop and FM spin-flop phases.
It is useful to recalling that $\Delta_3 = \lvert\lambda^*_0(\mathbf{k})\lambda_1(\mathbf{k})-\lambda_0(\mathbf{k})\lambda_1(-\mathbf{k})\rvert^2$,
it is easy to magnify $\Delta_3$ as $\Delta_3 \leq |\lambda_0(\mathbf{k})|^2(|\lambda_1(\mathbf{k})|+|\lambda_1(-\mathbf{k})|)^2$
using the triangle inequality theorem.
Since $|\varepsilon_0| = 2|\tilde{\Gamma}| \neq 0$,
the goal $\Delta_3 < \Delta_1$ can be achieved when a stronger relation of $\Pi(\psi,h,\varphi;\mathbf{k}) < 1$ is established, where
\begin{equation}
\Pi(\psi,h,\varphi;\mathbf{k}) = \frac{|\lambda_1(-\mathbf{k})|+|\lambda_1(\mathbf{k})|}{2|\varepsilon_0|}.
\end{equation}
Here, expression for $\lambda_1(\mathbf{k})$ can be found in Eq.~\eqref{EQ:AFMFldALmabda1} (for AFM spin-flop phase) and Eq.~\eqref{EQ:FMFldALmabda1} (for FM spin-flop phase) in the main text.
For convenience, we repeat them as follows:
\begin{equation}\label{EQSM:Lambda1SpinFlop}
\lambda_1(\mathbf{k})=\left\{
\begin{array}{lcl}
\frac{1}{6}(c_4f_{\mathbf{k}}^*+c_5+c_6ig_{\mathbf{k}}^*),  &      & {\textrm{AFM spin-flop phase}}\\
\frac{1}{6}\Big\{c_4f_{\mathbf{k}}^*+c_5+c_6g_{\mathbf{k}}^*+2i\big[c_7(f_{\mathbf{k}}^*-3)+c_8g_{\mathbf{k}}^*\big]\Big\},  &      & {\textrm{FM spin-flop phase}} \\
\end{array} \right..
\end{equation}
Due to the complexity of $\Pi$, it is challenging to calculate analytically the maximal value of $\Pi$ throughout the parameter space of spin-flop phases.
We therefore resort to numerical calculation to aid our proof.

\begin{figure}[htbp]
  \centering
  \includegraphics[width=0.96\linewidth]{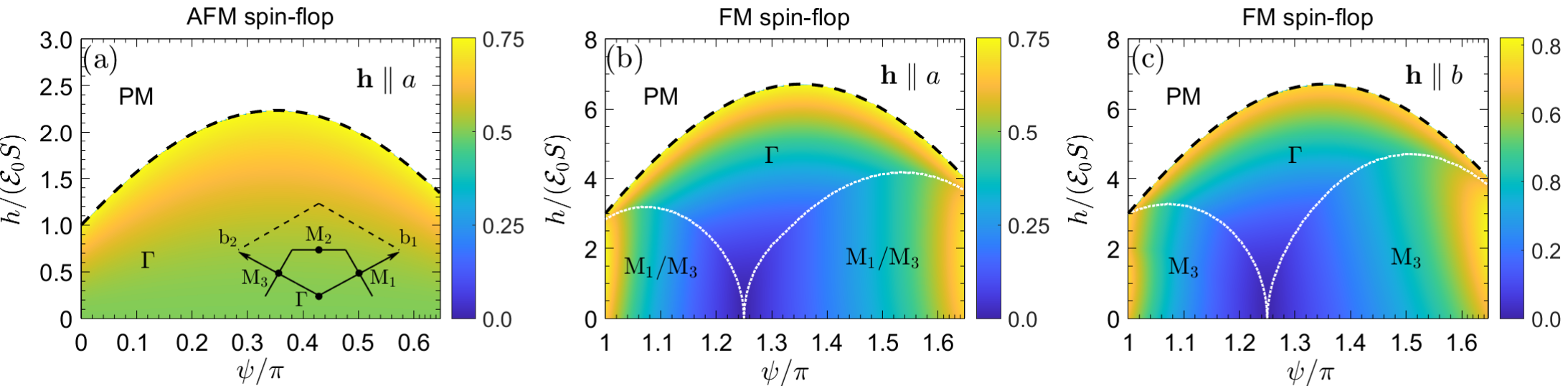}
  \caption{The landscape of function $\max_{\mathbf{k}}[\Pi(\psi,h,\varphi;\mathbf{k})]$ in the spin-flop phases under an in-plane magnetic field.
  (a) AFM spin-flop phase under the $a$-directional magnetic field; (b,c) FM spin-flop phase under the $a$-directional and $b$-directional magnetic fields, respectively.
  The symbols $\boldsymbol{\Gamma}$, $\textbf{M}_1$, and $\textbf{M}_3$ mark the region
  where function $\Pi(\psi,h,\varphi;\mathbf{k})$ obtains its maximal value in the reciprocal space.}
\label{FigSM:Supp-PiFunction}
\end{figure}

For the AFM spin-flop phase, $\Pi$ always reaches its maximum value at the $\boldsymbol{\Gamma}(0, 0)$ point despite the field direction.
When $\mathbf{k} = \boldsymbol{\Gamma}$, $f_\mathbf{k}=3$, $g_\mathbf{k}=0$, the expression for $\Pi$ is given by
\begin{equation}
\Pi(\psi,h,\varphi;\boldsymbol{\Gamma}) = 0.5 + \frac{h^2}{4h_c^2},
\end{equation}
which reaches its maximum value of 0.75 when $h = h_c$. The landscape of $\max_{\mathbf{k}}[\Pi(\psi,h,\varphi;\mathbf{k})]$ is shown in Fig.~\ref{FigSM:Supp-PiFunction}(a).

For the FM spin-flop phase, $\Pi$ can reach its maximum value at either $\boldsymbol{\Gamma}$ point or $\textbf{M}_1$, $\textbf{M}_3$ points.
At these points, the condition $|\lambda_1(\mathbf{k})|=|\lambda_1(-\mathbf{k})|$ is satisfied, and $\Pi$ can be reexpressed as $\Pi = |\lambda_1(\mathbf{k})|/\varepsilon_0$.
Discussion of $\Pi$ at these high symmetry points is enumerated as follows.

\begin{itemize}
  \item When $\mathbf{k} = \boldsymbol{\Gamma}$, $f_\mathbf{k}=3$, $g_\mathbf{k}=0$,
    \begin{align}
    \Pi(\psi,h,\varphi;\boldsymbol{\Gamma})= \frac{3}{4}\left(\frac{h}{h_c}\right)^2.
    \end{align}
  \item When $\mathbf{k} = \textbf{M}_1$, $f_\mathbf{k}=1$, $g_\mathbf{k}=-2$,
    \begin{align}
    \Pi^2(\psi,h,\varphi;\textbf{M}_1)=\frac{1}{(12\widetilde\Gamma)^2}\bigg\{
    &\Big[3\widetilde\Gamma\sin^2\theta-2\overline\Gamma\big(2\sin(2\varphi+\pi/6)(\sin^2\theta-2)+\sqrt{2}\sin2\theta\sin(\varphi-\pi/6)\big)\Big]^2 \nonumber \\
    &+16\overline\Gamma^2\big(2\cos\theta\sin(2\varphi-\pi/3)-\sqrt{2}\sin\theta\sin(\varphi+\pi/3)\big)^2 \bigg\}.
    \end{align}
  \item When $\mathbf{k} = \textbf{M}_3$, $f_\mathbf{k}=1$, $g_\mathbf{k}=2$,
    \begin{align}
    \Pi^2(\psi,h,\varphi;\textbf{M}_3)=\frac{1}{(12\widetilde\Gamma)^2}\bigg\{
    &\Big[3\widetilde\Gamma\sin^2\theta+2\overline\Gamma\big(2\sin(2\varphi-\pi/6)(\sin^2\theta-2)-\sqrt{2}\sin2\theta\sin(\varphi+\pi/6)\big)\Big]^2 \nonumber \\
    &+16\overline\Gamma^2\big(2\cos\theta\sin(2\varphi+\pi/3)-\sqrt{2}\sin\theta\sin(\varphi-\pi/3)\big)^2 \bigg\}.
    \end{align}
\end{itemize}

Specifically, when the field is along the $a$ direction, the expression of $\Pi(\textbf{M}_1/\textbf{M}_3,\varphi=0)$ is
\begin{align}
\Pi(\psi,h,\varphi = 0;\textbf{M}_1/\textbf{M}_3)&=\frac{\sqrt{(c_4+c_5)^2+16c_8^2}}{6\varepsilon_0}\\
   &=\frac{\sqrt{\big[3\widetilde\Gamma\sin^2\theta-\overline\Gamma(2\sin^2\theta+\sqrt{2}\sin2\theta-4)\big]^2+24\overline\Gamma^2(\sqrt{2}\cos\theta+\sin\theta)^2}}{-12\widetilde\Gamma}.
\end{align}
When the field is along the $b$ direction, the expression of $\Pi(\textbf{M}_3,\varphi=\pi/2)$ is
\begin{align}
\Pi(\psi,h,\varphi = \pi/2;\textbf{M}_3)&=\frac{\sqrt{(c_4+c_5+2c_6)^2+16(c_8-c_7)^2}}{6\varepsilon_0}\\
   &=\frac{\sqrt{\big[3\widetilde\Gamma\sin^2\theta+\overline\Gamma(2\sin^2\theta-\sqrt{6}\sin2\theta-4)\big]^2+8\overline\Gamma^2(\sqrt{6}\cos\theta+\sin\theta)^2}}{-12\widetilde\Gamma}.
\end{align}
The landscape of $\max_{\mathbf{k}}[\Pi(\psi,h,\varphi;\mathbf{k})]$ for $\varphi = 0$ and $\pi/2$
are shown in Fig.~\ref{FigSM:Supp-PiFunction}(b) and Fig.~\ref{FigSM:Supp-PiFunction}(c), respectively.

To conclude, given that $\Pi(\psi,h,\varphi;\mathbf{k})$ is always less than 1,
it can be concluded that $\Delta_3<\Delta_1$ holds in both the AFM and FM spin-flop phases.

\begin{figure}[!ht]
    \centering
    \includegraphics[width=1.0\columnwidth]{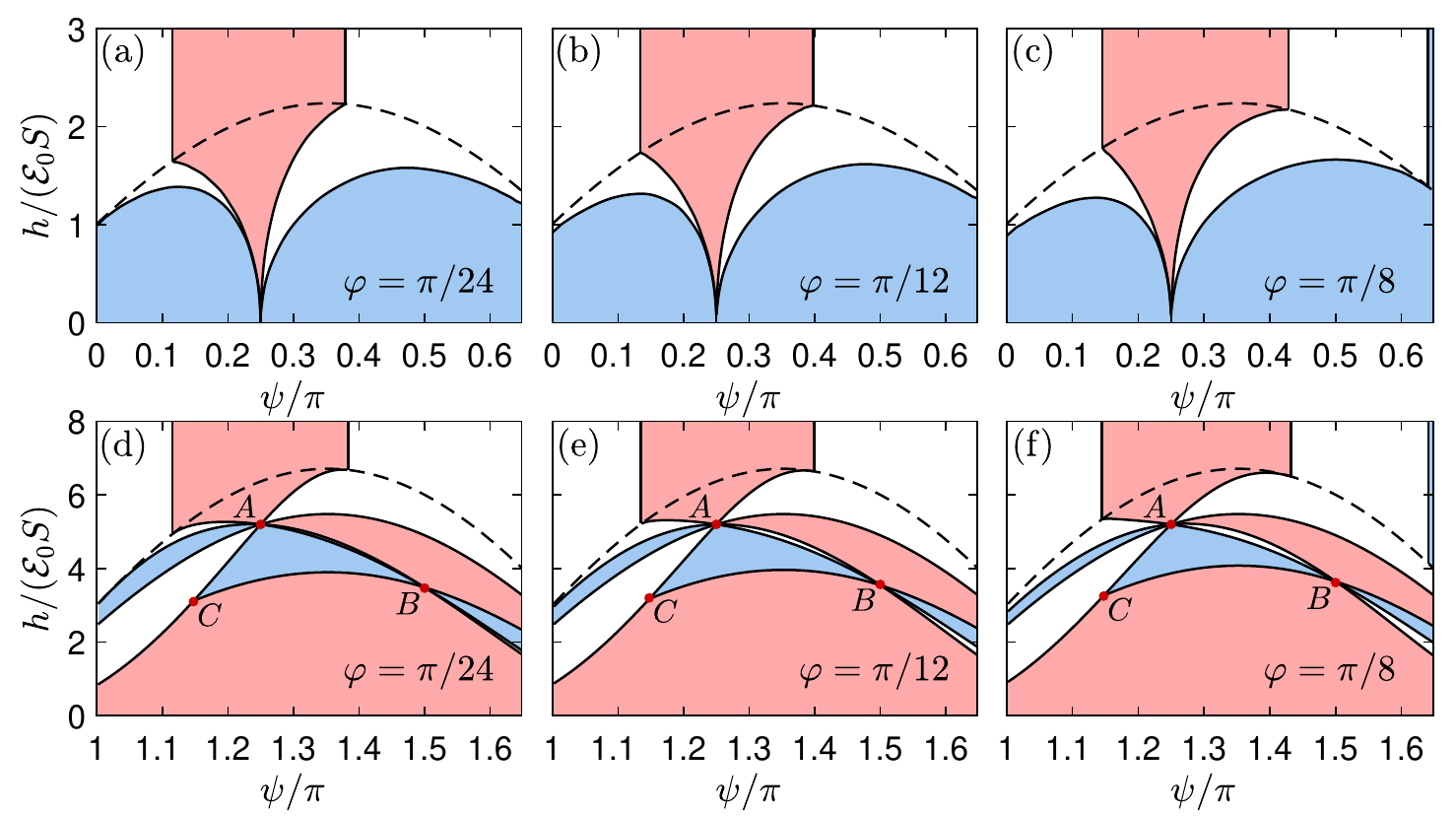}
    \caption{Topological phase diagrams of (a-c) the AFM spin-flop phase and (d-f) the FM spin-flop phase at $\varphi$ equal to (a, d) $\pi/24$, (b, e) $\pi/12$, and (c, f) $\pi/8$. Red, white, and blue areas indicate $\mathcal{C}_1$ = $+1$, $0$, and $-1$, respectively. Black solid (dashed) curve mark the topological (magnetic) phase transitions.}
    \label{fig:Supp-diagram}
\end{figure}

\section{Topological phase diagrams in general directions}\label{SecSM4}

Here we show the topological phase diagrams of the AFM spin-flop and the FM spin-flop phases at the in-plane field angles $\varphi$ = $\pi/24$, $\pi/12$, and $\pi/8$ in Fig.~\ref{fig:Supp-diagram}, and the $\varphi$-dependence of the animation with one-degree increments is relegated to Video1 and Video2, respectively.
In the paramagnet at high field, flipping the signs of all couplings, i.e., $\psi\rightarrow\psi+\pi$, the Chern number $\mathcal{C}_n$ remains invariant \cite{SMChern2024PRB}.
Therefore, it is interesting to note that patterns of the Chern number in the paramagnet at panels (a) and (d) are the same.
The same conclusion can be drawn for panels (b) and (e), as well as for panels (c) and (f).

\section{Chiral Edge State in FM spin-flop phase}\label{SecSM5}
The nontrivial band topology can be confirmed by calculating the chiral edge states in a nanoribbon geometry.
When the open boundary condition is applied, there will be chiral edge modes connecting the upper and lower magnon bands.
Here, we take the FM spin-flop phase in the $a$-directional magnetic field as an example to calculate the chiral edge states in the zigzag nanoribbon geometry \cite{SMNunez2020PRB}.

\begin{figure}[!ht]
    \centering
    \includegraphics[width=0.7\linewidth]{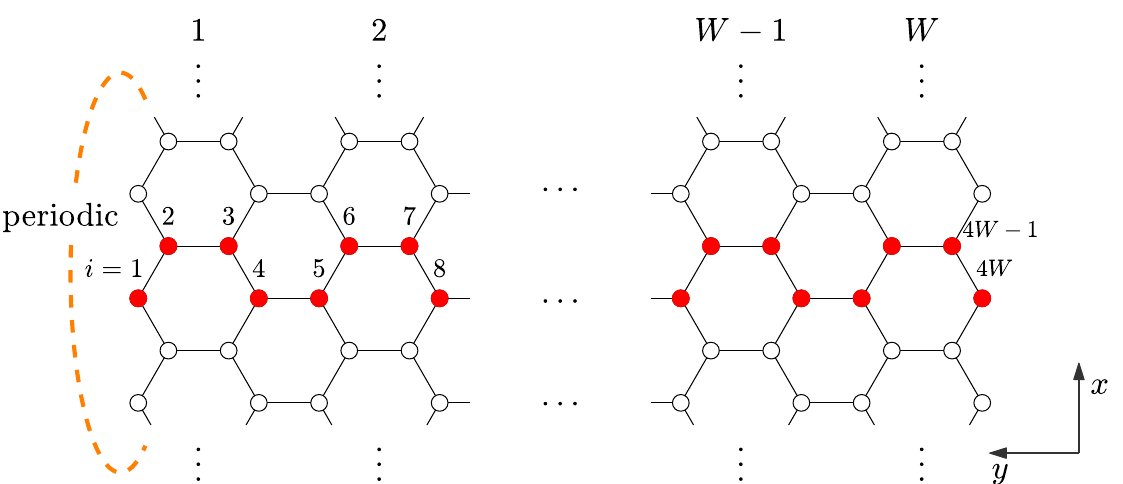}
    \caption{The honeycomb-lattice nanoribbon has open boundary conditions along the armchair direction and periodic boundary conditions along the zigzag direction.
    The nanoribbon consists of $W$ periodic one-dimensional chains, with the numbers near the sites representing the $i$ indices.}
\label{Fig:EdgeState}
\end{figure}

\begin{figure}[!ht]
    \centering
    \includegraphics[width=0.95\linewidth]{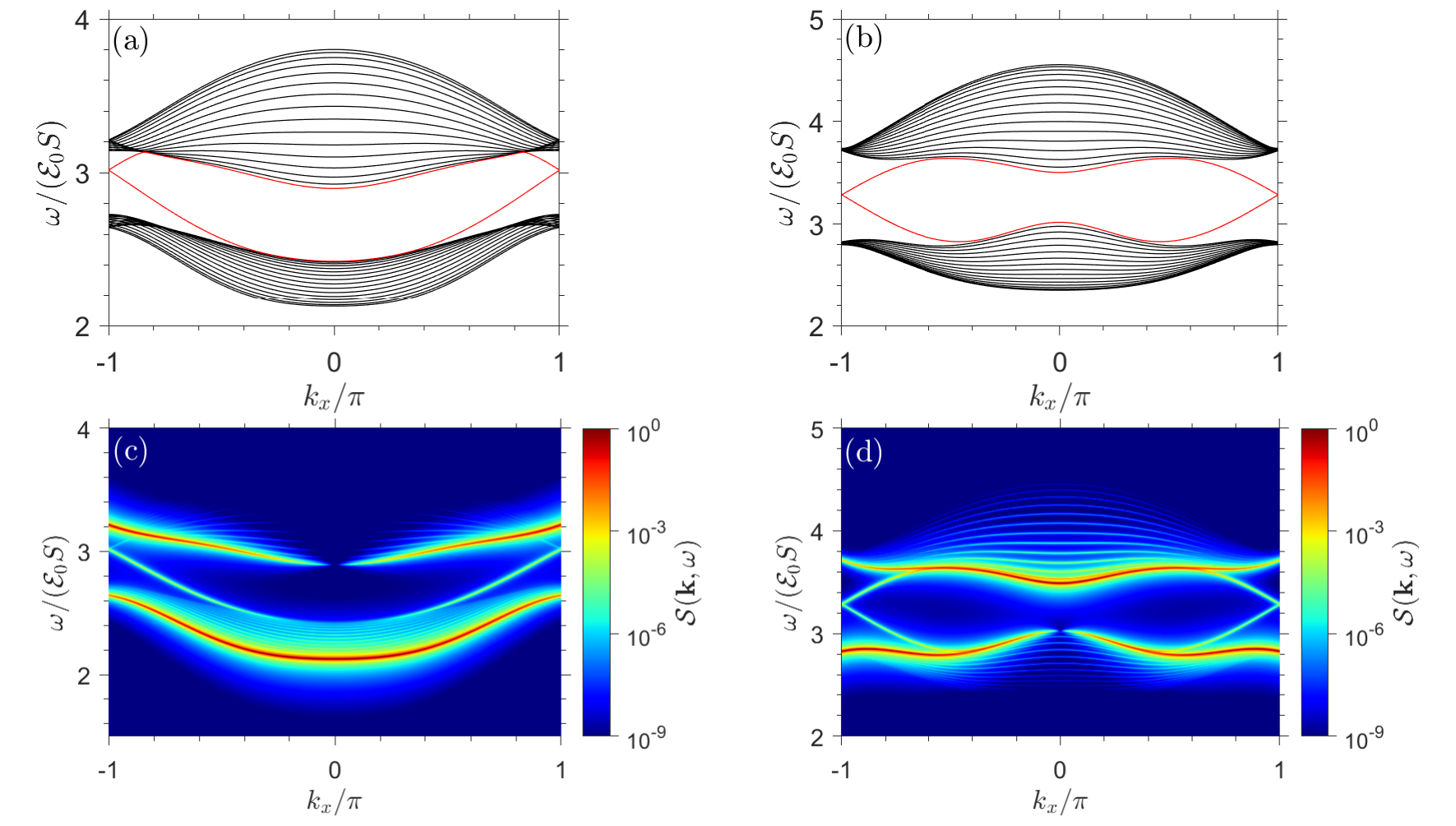}
    \caption{Magnon bands (a,b) and dynamical spin structure factors (c,d) in the zigzag nanoribbon geometry with $W=8$ at (a,c) $\psi/\pi = 1.10$ and (b,d) $\psi/\pi = 1.55$, the magnetic field $h/(\mathcal{E}_0S) = 4.2$ is applied along the $a$ direction. (a,b) Chiral edge modes are depicted by red lines. (c,d) In each case, the intensity is normalized such that max $\mathcal{S}(\mathbf{k},\omega)=1$ and the intensity scale is logarithmic from $1\times10^{-9}$ to $1$.}
\label{Fig:DSF}
\end{figure}

Since the momentum $\mathbf{k}$ in the $y$ direction is no longer a good quantum number, we rewrite the Hamiltonian in the ($k_x$, $y$) space (see Fig.~\ref{Fig:EdgeState}).
By performing a Fourier transformation, the linear spin wave Hamiltonian reads
\begin{align}\label{EQSM:LSWT}
    \mathcal{H}=\frac{S}{2}\sum_{\mathbf{k}}\mathrm{\Phi}^{\dagger}_\mathbf{k}\mathcal{H}_{\mathbf{k}}\mathrm{\Phi}_\mathbf{k}, ~~~~~~\mathcal{H}_\mathbf{k} =
	\left(\begin{array}{@{}cc}
	\mathbf{A_{k}}          &      \mathbf{B_{k}}         \\
    \mathbf{B}^{\dag}_{-\mathbf{k}}        &   \mathbf{A}^{\mathrm{T}}_{-\mathbf{k}}             \\
	\end{array}\right),
\end{align}
where $\mathrm{\Phi}_\mathbf{k}=(\alpha_{1,k_x},\alpha_{2,k_x},\dots,\alpha_{4W-1,k_x},\alpha_{4W,k_x},\alpha^{\dagger}_{1,-k_x},\alpha^{\dagger}_{2,-k_x},\dots,\alpha^{\dagger}_{4W-1,-k_x},\alpha^{\dagger}_{4W,-k_x})^{\mathrm{T}}$.
The $\mathbf{A_{k}}$ and $\mathbf{B_{k}}$, whose dimensions are $4W\times4W$, are given by
\begin{align}
\mathbf{A_{k}}=
\left[
\begin{array}{ccccc}
G(k_x) & F & 0 & \cdots & 0\\
F^{\dagger} & G(k_x) & F & \ddots & \vdots \\
0 & F^{\dagger} & \ddots & \ddots & 0 \\
\vdots & \ddots & \ddots & \ddots & F \\
0 & \cdots & 0 & F^{\dagger} & G(k_x) \\
\end{array}
\right],
\mathbf{B_{k}}=
\left[
\begin{array}{ccccc}
Q(k_x) & P & 0 & \cdots & 0\\
P^{\dag} & Q(k_x) & P & \ddots & \vdots \\
0 & P^{\dag} & \ddots & \ddots & 0 \\
\vdots & \ddots & \ddots & \ddots & P \\
0 & \cdots & 0 & P^{\dag} & Q(k_x) \\
\end{array}
\right],
\end{align}
where $G(k_x),F,Q(k_x)$ and $P$ are $2\times2$ matrices
\begin{align}
G(k_x)=
\left[
\begin{array}{cc}
G_{11} & G_{12}\\
G_{21} & G_{22} \\
\end{array}
\right],
F=
\left[
\begin{array}{cc}
0 & 0\\
F_{21} & 0 \\
\end{array}
\right],
Q(k_x)=
\left[
\begin{array}{cc}
Q_{11} & Q_{12}\\
Q_{21} & Q_{22} \\
\end{array}
\right],
P=
\left[
\begin{array}{cc}
0 & 0\\
P_{21} & 0 \nonumber
\end{array}
\right],
\end{align}
\begin{align}
&G_{11}=G_{22}=-2\widetilde\Gamma,\\\nonumber
&G_{12}(k_x)=G^*_{21}(k_x)=\frac{1}{3}\Big[\widetilde\Gamma(3\sin^2\theta-2)+\overline\Gamma(\sin^2\theta
      +\sqrt{2}\sin\theta\cos\theta)\Big]\cos\big(\frac{\sqrt{3}k_x}{2}\big),\\\nonumber
&F_{21}=\frac{1}{6}\Big[\widetilde\Gamma(3\sin^2\theta-2)-2\overline\Gamma(\sin^2\theta
      +\sqrt{2}\sin\theta\cos\theta)\Big],\\\nonumber
&Q_{12}(k_x)=Q^*_{21}(k_x)=\frac{1}{3}\Big\{3\widetilde\Gamma\sin^2\theta+\overline\Gamma\big[(\sin^2\theta-2) + \sqrt{2}\sin\theta\cos\theta \big]\Big\}\cos\big(\frac{\sqrt{3}k_x}{2}\big)+\frac{\sqrt{6}\,\overline\Gamma}{3}(\sin\theta+\sqrt{2}\cos\theta)\sin\big(\frac{\sqrt{3}k_x}{2}\big),\\\nonumber
&P_{21}=\frac{1}{6}\Big\{3\widetilde\Gamma\sin^2\theta - 2\overline\Gamma\big[(\sin^2\theta-2)+\sqrt{2}\sin\theta\cos\theta \big]\Big\}.
\end{align}
Then, the $\mathcal{H}_\mathbf{k}$ is diagonalized by the Bogoliubov transformation
\begin{align}\label{EQSM:Bogoliubov}
\left(\begin{array}{@{}cc@{}}
	E_{\mathbf{k}} & 0 \\
	0 & -E_{\mathbf{k}}
	\end{array}\right)=\mathcal{T}_{\mathbf{k}}^{\dagger}\left(\begin{array}{@{}cc}
	\mathbf{A_{k}}          &      \mathbf{B_{k}}         \\
    \mathbf{B}^{\dag}_{-\mathbf{k}}        &   \mathbf{A}^{\mathrm{T}}_{-\mathbf{k}}             \\
	\end{array}\right)\mathcal{T}_{\mathbf{k}},
    ~~~~~~~~
	\mathcal{T}_{\mathbf{k}} =
	\left(\begin{array}{@{}cc@{}}
	U_{\mathbf{k}} & V_{\mathbf{k}} \\
	V^*_{\mathbf{-k}} & U^*_{\mathbf{-k}}
	\end{array}\right).
\end{align}
where $E_{\mathbf{k}}=\mathrm{diag}(\omega_{1,k_x},\omega_{2,k_x},\dots,\omega_{4W-1,k_x},\omega_{4W,k_x})^{\mathrm{T}}$.
Meanwhile, the transformation matrix $\mathcal{T}_{\mathbf{k}}$ satisfies the orthogonality relations $\boldsymbol{\Sigma} = \mathcal{T}_{\mathbf{k}}^{\dagger}\boldsymbol{\Sigma} \mathcal{T}_{\mathbf{k}}$
with $\boldsymbol{\Sigma} = \mathrm{diag}(\mathbb{1}, \mathbb{1}, \mathbb{-1}, \mathbb{-1})$, where $\mathbb{1}$ is $2W\times2W$ identity matrix.
The probability density of the wave function is expressed as \cite{SMDiaz2020PRR}
\begin{align}\label{EQ:WaveFunction}
|\Psi^{(n)}_{(i,k_x)}|^2=|u_{(i,k_x)}^{(n)}|^2+|v_{(i,k_x)}^{(n)}|^2,
\end{align}
where $i$ label is the spin sites (see Fig.~\ref{Fig:EdgeState}) and $u_{i,k_x}^{(n)}$ and $v_{i,k_x}^{(n)}$ are the $(i,n)$ element of $U_{\mathbf{k}}$ and $V_{\mathbf{k}}$, respectively.
By Eq.~\eqref{EQSM:Bogoliubov} and Eq.~\eqref{EQ:WaveFunction}, we obtain the magnon bands and probability density of the wave function at the $i$ site, respectively.
We choose $W = 8$ to ensure that the results converge with $W$ and the result is shown in Fig.~\ref{Fig:DSF}(a,b) and Fig. 8 in the main text.

\section{Dynamic Spin Structure Factor in FM spin-flop phase}\label{SecSM6}

The dynamic spin structure factor $\mathcal{S}(\mathbf{k},\omega)$ reflects the time-dependent spin correlations and is in principle experimentally accessible by inelastic neutron or X-ray scattering. The dynamical spin structure factor is defined as
\begin{equation}
  \mathcal{S}(\mathbf{k},\omega)=\int \mathrm{d}t~e^{i\omega t}\sum_{ij}\big \langle \mathbf{S}_i(t)\cdot \mathbf{S}_j(0)\big \rangle~e^{i\mathbf{k}\cdot(\mathbf{r}_j-\mathbf{r}_i)}.
\end{equation}
In linear spin-wave theory, when the open boundary condition is applied, we can further calculate the dynamical spin structure factor using \cite{SMJoshi2018PRB,SMVojta2019JPCM}
\begin{equation}
\mathcal{S}(\mathbf{k},\omega)=\frac{S}{2} \sum_{n=1}^{4W} \sum_{m,m'=1}^{4W} 2\pi \delta(\omega-\omega_{n,k_x})\left[ u^{(n)}_{k_xm} u_{k_x m'}^{*(n)}  + v_{-k_xm}^{*(n)} v_{k_x m'}^{(n)} \right] + \mathcal O(\delta(\omega), S^0),
\end{equation}
where $u_{k_xm}^{(n)}$ and $v_{-k_xm}^{*(n)}$ are the $(m,n)$ element of $U_{\mathbf{k}}$ and $V^*_{-\mathbf{k}}$ by Eq.~\eqref{EQSM:Bogoliubov}, respectively.
Here, we calculate the dynamical spin structure factor at $\psi/\pi = 1.10$ and $\psi/\pi = 1.55$ in the $a$-directional magnetic field $h/(\mathcal{E}_0S) = 4.2$ in Figs.~\ref{Fig:DSF}(c,d).
As in Figs.~\ref{Fig:DSF}(a,b), distinct edge modes appear between the bulk modes, with the edge modes interacting at $k_x = \pm\pi$.
As mentioned in the main text, this provides further evidence for the existence of topological magnons.

%


\end{document}